\documentclass[a4paper,11pt]{article}
\usepackage[T1]{fontenc}

\usepackage{xcolor}
\definecolor{myblue}{HTML}{0084ff}

\usepackage[pdftex,colorlinks]{hyperref}
\hypersetup{colorlinks,%
		    citecolor=myblue,%
		    filecolor=myblue,%
		    linkcolor=myblue,%
		    urlcolor=myblue}

\usepackage[left=1.1in,right=1.1in,bottom=1.5in,top=1.2in]{geometry}
\usepackage{txfonts}

\newcommand{\newresults}{
\makeatletter
\if@twocolumn%
	\backgroundsetup{
		position={0.572\paperwidth,-0.025\paperheight} ,
		angle=270,
		color=black,
		opacity=0.60,
		scale=1.5,
		contents={\tikz\node[text=black,fill=gray!40,align=left, minimum width=2.3cm,minimum height=0.6cm,inner sep=0]{New Results};}}
\else
	\backgroundsetup{
		position={0.338\paperwidth,-0.06\paperheight} ,
		angle=270,
		color=black,
		opacity=0.60,
		scale=2.5,
		contents={\tikz\node[text=black,fill=gray!40,align=left, minimum width=2.3cm,minimum height=0.6cm,inner sep=0]{New Results};}}
\fi%
}

\newcommand{\confirmatoryresults}{
\makeatletter
\if@twocolumn%
\backgroundsetup{
	position={0.572\paperwidth,-0.03\paperheight} ,
	angle=270,
	color=black,
	opacity=0.60,
	scale=1.5,
	contents={\tikz\node[text=black,fill=gray!40,align=left, minimum width=3.9cm,minimum height=0.6cm,inner sep=0]{Confirmatory Results};}}
\else
	\backgroundsetup{
	position={0.338\paperwidth,-0.07\paperheight} ,
	angle=270,
	color=black,
	opacity=0.60,
	scale=2.5,
	contents={\tikz\node[text=black,fill=gray!40,align=left, minimum width=3.9cm,minimum height=0.6cm,inner sep=0]{Confirmatory Results};}}
\fi%
}

\newcommand{\contradictoryresults}{
\makeatletter
\if@twocolumn%
	\backgroundsetup{
	position={0.572\paperwidth,-0.03\paperheight} ,
	angle=270,
	color=black,
	opacity=0.60,
	scale=1.5,
	contents={\tikz\node[text=black,fill=gray!40,align=left, minimum width=4.0cm,minimum height=0.6cm,inner sep=0]{Contradictory Results};}}
\else
	\backgroundsetup{
	position={0.338\paperwidth,-0.07\paperheight} ,
	angle=270,
	color=black,
	opacity=0.60,
	scale=2.5,
	contents={\tikz\node[text=black,fill=gray!40,align=left, minimum width=4.0cm,minimum height=0.6cm,inner sep=0]{Contradictory Results};}}
\fi%
}

\usepackage{float}

\usepackage{rotating}

\usepackage{supertabular}

\usepackage{caption}

\usepackage{authblk}

\makeatletter
\renewcommand{\maketitle}{\bgroup\setlength{\parindent}{0pt}
\begin{flushleft}
  \thispagestyle{plain}
  \textbf{\@title}

  \@author
\end{flushleft}\egroup
}
\makeatother

\renewenvironment{abstract}
 {\small
  \begin{flushleft}
  \large \textbf{\abstractname}\vspace{-0.40em}\vspace{0pt}
  \end{flushleft}
  \list{}{
    \setlength{\leftmargin}{0cm}%
    \setlength{\rightmargin}{\leftmargin}%
  }%
  \item\relax}
 {\endlist}

\usepackage{amssymb}
\usepackage{amsfonts}
\usepackage{lscape}
\usepackage{booktabs}

\usepackage{makecell}
\usepackage{xcolor}
\usepackage{graphicx}
\usepackage{dcolumn}
\usepackage{bm}
\usepackage{hyperref}

\newcolumntype{R}[1]{>{\hfill}p{#1}}

\makeatletter
\newcommand{\algrule}[1][.2pt]{\par\vskip.5\baselineskip\hrule height #1\par\vskip.5\baselineskip}
\makeatother

\usepackage[sorting=none,
            style=numeric-comp,
            bibstyle=ieee,
            url=false,
            doi=false,
            isbn=false,
            language=english,
            clearlang=true,
            maxbibnames=5,
            giveninits=true]{biblatex}

\DeclareFieldFormat[article]
{volume}{\textbf{#1}} 
\DeclareFieldFormat[article, inbook, incollection, inproceedings, misc, thesis, unpublished]
{number}{(#1)}
\DeclareFieldFormat[article,inproceedings]
{pages}{#1}
\DeclareFieldFormat[inbook, incollection]
{pages}{p. #1}
\renewbibmacro*{volume+number+eid}{
  \printfield{volume}\printfield{number}:
}

\AtEveryBibitem{\clearfield{month}}
\AtEveryBibitem{\clearfield{day}}
\AtEveryBibitem{\clearfield{note}}

\AtEveryBibitem{\clearlist{language}}

\newcommand{\given}{{\, | \,}}
\newcommand{\mailto}[1]{\href{mailto:#1}{#1}}

\newcommand{\dif}{}
\usepackage{mathtools}

\usepackage{mhchem}

\newcommand{\singlereac}[1]{\xrightarrow{\hspace{1ex}#1\hspace{1ex}}}
\newcommand{\doublereac}[2]{\ce{<=>[{#1}][{#2}]}}

\newcommand{\E}{\mathbb{E}}

\renewcommand{\dif}{\mathrm{d}}

\DeclareMathOperator{\KL}{\mathrm{KL}}

\newcommand{\bn}{{\bm{n}}}
\newcommand{\bnp}{{\bm{m}}}
\newcommand{\bz}{{\bm{z}}}

\usepackage[labelfont=bf]{caption}

\bibliography{references}

\usepackage{algorithm}
\usepackage[noend]{algpseudocode}

\title{\LARGE Model Reduction for the Chemical Master Equation: \newline an  Information-Theoretic Approach \newline}

\author[1,2]{Kaan \"Ocal}
\author[3]{Guido Sanguinetti}
\author[2,$\dagger$]{Ramon Grima}
\affil[1]{School of Informatics, University of Edinburgh, Edinburgh EH8 9AB, UK}
\affil[2]{School of Biological Sciences, University of Edinburgh, Edinburgh EH9 3JH, UK}
\affil[3]{Scuola Internazionale Superiore di Studi Avanzati, 34136 Trieste, Italy}
\affil[$\dagger$]{Corresponding author: \mailto{ramon.grima@ed.ac.uk}}

\date{}

\begin{document}

\maketitle
\begin{abstract}
The complexity of mathematical models in biology has rendered model reduction an essential tool in the quantitative biologist's toolkit. For stochastic reaction networks described using the Chemical Master Equation, commonly used methods include time-scale separation, the Linear Mapping Approximation and state-space lumping. Despite the success of these techniques, they appear to be rather disparate and at present no general-purpose approach to model reduction for stochastic reaction networks is known. In this paper we show that most common model reduction approaches for the Chemical Master Equation can be seen as minimising a well-known information-theoretic quantity between the full model and its reduction, the Kullback-Leibler divergence defined on the space of trajectories. This allows us to recast the task of model reduction as a variational problem that can be tackled using standard numerical optimisation approaches. In addition we derive general expressions for the propensities of a reduced system that generalise those found using classical methods. We show that the Kullback-Leibler divergence is a useful metric to assess model discrepancy and to compare different model reduction techniques using three examples from the literature: an autoregulatory feedback loop, the Michaelis-Menten enzyme system and a genetic oscillator.
\end{abstract}


\section{Introduction}


Stochastic biochemical reaction networks such as those involved in gene expression, immune response or cellular signalling \cite{kaern_stochasticity_2005,satija_heterogeneity_2014,lipniacki_stochastic_2008,schnoerr_approximation_2017} are often described using the Chemical Master Equation (CME). The CME describes the dynamics of biochemical processes on a mesoscopic level, viewing them as a discrete collection of molecules interacting and undergoing reactions stochastically; as such it is generally considered more accurate than continuum approximations such as rate equations and the Chemical Langevin Equation \cite{schnoerr_approximation_2017}. Despite its explanatory power, the CME poses significant analytical and computational difficulties to modellers that have limited its use in practice. Closed-form solutions to the CME are difficult to obtain and are only known for a small number of biologically relevant systems, and solving the CME numerically requires using approximations such as the Finite State Projection (FSP) \cite{munsky_finite_2006}. Numerical approaches tend to scale poorly with the number of species and reactions present in a system, and as a result there is significant interest in finding ways to simplify a description of a stochastic reaction network that make it easier to analyse and study - this is the goal of model reduction.

Model reduction for deterministic and continuum-limit models in biology is an active research topic \cite{ali_eshtewy_model_2020,snowden_methods_2017}, but very few existing methods can be applied to the discrete, stochastic setting of the CME. The Quasi-Steady State Approximation (QSSA) is perhaps the best known technique, first considered in the stochastic case in \cite{rao_stochastic_2003}. Here the system is partitioned into `slow` and `fast` species such that the fast species evolve very quickly on the timescale of the slow species. On the slow timescale the states of the fast species can therefore be approximated by their steady-state value (conditioned on the slow species), effectively allowing a description of the system in terms of the slow species only. By its nature the QSSA is only applicable to systems with a clear separation of timescales between species, the existence of which cannot always be established. The QSSA for stochastic systems generally requires more stringent conditions than in the deterministic case, but the exact validity conditions are not well-understood \cite{kim_relationship_2015,kim_validity_2014,thomas_slow-scale_2012,thomas_communication_2011,kang_quasi-steady-state_2019,eilertsen_stochastic_2022}.

Similar to the QSSA is the Quasiequilibrium Approximation (QEA), which was first considered in \cite{haseltine_approximate_2002,goutsias_quasiequilibrium_2005} for stochastic reaction networks. Here the reaction network is decomposed into `slow' and `fast' \emph{reactions}, and the fast reactions are assumed to equilibrate rapidly on the timescale of the slow reactions. Similar to the QSSA, the QEA can be used to reduce the number of species and reactions in a system, but it relies on the existence of a clear timescale separation between reactions, which is not always present for large systems with many distinct reactions. Much like the QSSA, the validity of the QEA for systems without the appropriate timescale separation has not been generally established, and from the asymptotic nature of the descriptions it is not usually possible to quantify the approximation error. Despite this, both the QSSA and the QEA are by far the most commonly used model-reduction technique for chemical reaction networks owing to their physical interpretability and analytical tractability, most famously in the Michaelis-Menten model of enzyme kinetics.


A distinct approach for model reduction with the Chemical Master Equation  is state-space lumping, which originates from the theory of finite Markov chains, see e.g.~\cite{cardelli_exact_2021}. Here different states in a system are lumped together such that the coarse-grained system is still Markovian and can be modelled using the CME. For a typical biochemical reaction network it may not be possible to perform any nontrivial lumping while preserving Markovianity, whence approximate lumping methods have been considered e.g.~in \cite{amjad_generalized_2020,deng_optimal_2011,geiger_optimal_2015}. Here the coarse-grained system is approximated by a Markov process, and the approximation error quantified using the KL divergence between the original model and a lift of the approximation to the state space of the original model. State-space lumping for the CME often occurs in the context of the QEA, as states related by fast reactions are effectively lumped together, or averaged \cite{bo_multiple-scale_2017,holehouse_stochastic_2020,jia_dynamical_2020}. For this reason we will not consider this approach separately, although many of our considerations, such as the optimal form of the lumped propensity functions, extend to state-space lumping.

Finally, a more recent model reduction technique specifically for gene expression systems is the Linear Mapping Approximation (LMA) \cite{cao_linear_2018}. The LMA replaces bimolecular reactions of the form $G + P \rightarrow (\ldots)$, where $G$ is a binary species such as a gene, by a linear approximation where $P$ is replaced by its mean conditioned on $[G] = 1$. While the LMA does not reduce the number of species or reactions, reaction networks with linear reactions are generally easier to analyse: their moments can be computed exactly, and closed-form solutions are known for many cases \cite{jahnke_solving_2007,gadgil_stochastic_2005,zhou_analytical_2012,li_steady-state_2021}.

At a first glance these approaches - timescale separation, state space lumping and the Linear Mapping Approximation - seem rather disparate, and it is unclear what, if any, relationships there are between these approaches. In this paper we show that all of these methods can be viewed as minimising a natural information theoretic quantity, the Kullback-Leibler (KL) divergence, between the full and reduced models. In particular they can be seen as maximal likelihood approximations of the full model, and one can assess the quality of the approximation in terms of the resulting KL divergence. Based on these results we show how the KL divergence can be estimated and minimised numerically, therefore providing an automated method to choose between different candidate reductions of a given model in situations where none of the above model reduction techniques are classically applicable.

The KL divergences we consider in this paper are computed on the space of trajectories, and as such include both static information and dynamical information, in contrast to purely distribution-matching approaches. The KL divergence and similar information-theoretic measures between continuous-time Markov chains have previously been considered in \cite{opper_variational_2007,wildner_moment-based_2019} in the context of variational inference (with the true model and the approximation reversed compared to our approach), in \cite{bronstein_marginal_2018} to obtain approximate non-Markovian reductions, in \cite{moor_dynamic_2022} to analyse information flow for stochastic reaction networks and in \cite{khudabukhsh_approximate_2019} in quantifying model discrepancy for Markovian agent-based models. 

In Section 2 we introduce the mathematical framework in which we consider model reduction for the Chemical Master Equation, based on KL divergences between continuous-time Markov chains. We show how the KL divergence can be minimised analytically in some important cases, recovering standard results in the literature and providing a mathematical justification for commonly used mean-field arguments as in the QSSA, the QEA and the LMA. We furthermore provide numerical algorithms for estimating as well as minimising the KL divergence in cases where analytical solutions are not available. In Section 3 we illustrate the use of KL divergences as a metric for approximation quality using three biologically relevant examples: an autoregulatory feedback loop exhibiting critical behaviour, Michaelis-Menten enzyme kinematics, where we reanalyse the QSSA and the QEA, and an oscillatory genetic feedback system taken from \cite{thomas_slow-scale_2012}, for which we compare different reductions using our approach. Finally in Section 4 we discuss our observations and how our approach could be used as a stepping-stone towards automated reduction of complex biochemical reaction pathways.

\section{Methods}

\subsection{Stochastic Reaction Networks}

\begin{figure}[t]
    \centering
    \includegraphics{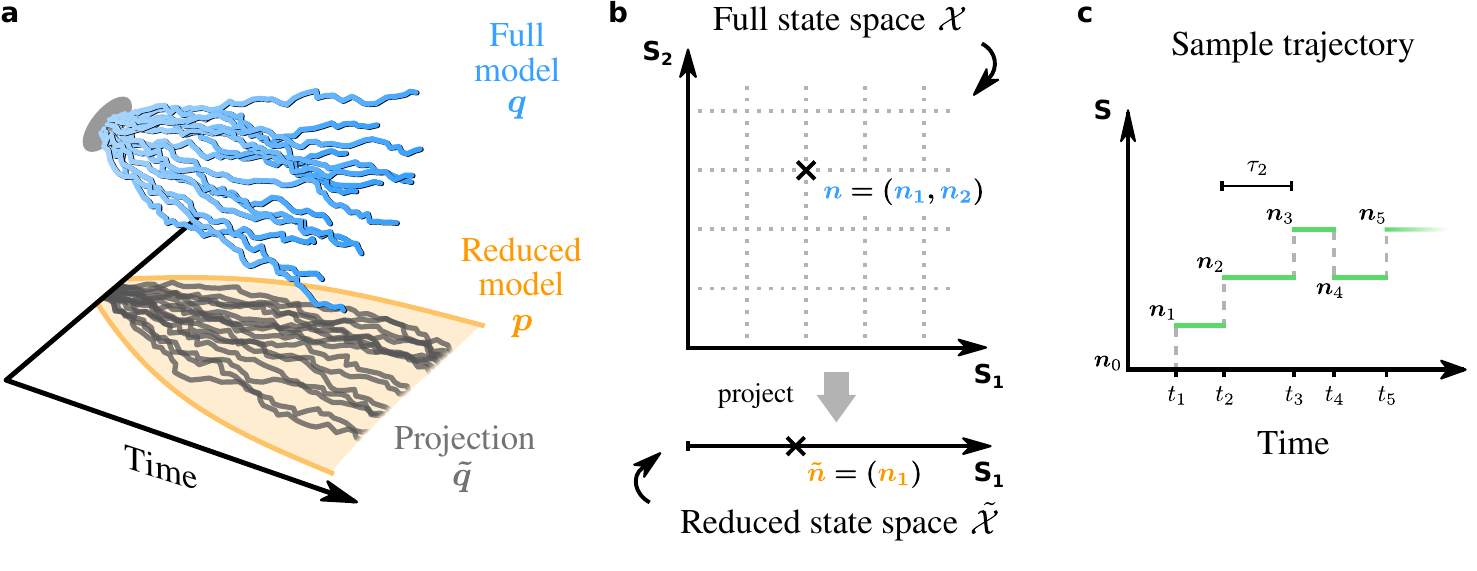}
    \caption{Model reduction for the Chemical Master Equation. \textbf{(a)} Model reduction approximates a high-dimensional model $q$ by a lower-dimensional version. Since the direct projection $\tilde q$ of the full model is not easy to describe, we approximate it using a family of tractable candidate models: in this paper, the approximation $p$ is described by the CME. \textbf{(b)} Comparison of the full state space of a system, consisting of two species $S_1$ and $S_2$, and a reduced state space containing $S_1$ only. Species which are not deemed essential can be discarded in the reduction and become unobserved variables. The dynamics of the original system involves all species, whereas the reduced model aims at an effective description only in terms of the reduced species. \textbf{(c)} Sample trajectory for a one-dimensional system, defined by the sequence $\bn_0, \bn_1, \ldots$ of states visited and the corresponding jump times $t_1$, $t_2$, ... (or alternatively the waiting times $\tau_0$, $\tau_1$, ...).}
    \label{fig:methods}
\end{figure}

The Chemical Master Equation describes a biochemical reaction network as a stochastic process on a discrete state space $\mathcal X$. We will use the letter $q$ to denote such a stochastic process, which for the purposes of this paper can be seen as a probability distribution over \emph{trajectories} on $\mathcal X$. For a biochemical reaction network the state space will be $\mathcal X = \mathbb{N}^s$, where $s$ is the number of species in the system: every state consist of a tuple $\bn = (n_1, \ldots, n_s)$ of $s$ integers describing the abundances of each species.

Since the model $q$ can consist of many species interacting in complicated ways, we often want to find a reduction that is more tractable, yet approximates $q$ as closely as possible. The reduced model, which we will call $p$, should be of the same form as $q$, i.e.~described by the CME, but will typically involve fewer species and simpler reactions. In particular $p$ can be defined on a lower-dimensional state space $\mathcal {\tilde X}$. A state $\bn$ in the original model can then be described by its projection $\tilde \bn$ onto this lower-dimensional space, together with some unobserved components, which we will denote $\bz$. See Fig.~\ref{fig:methods}\textbf{a}~and~\textbf{b} for an illustration. 

In this paper we assume that the basic structure of $p$ is known \emph{a priori}, i.e.~the species and reactions we wish to retain are fixed. Our approach to model reduction therefore consists in finding the optimal propensity functions for the reduced model, and we shall see how this can give rise to various known approximations such as the QSSA, the QEA or the LMA depending on what reductions are performed. We will return to the related problem of choosing the structure of the reduced model $p$ in the discussion.

\begin{figure}[t]
    \centering
    \includegraphics{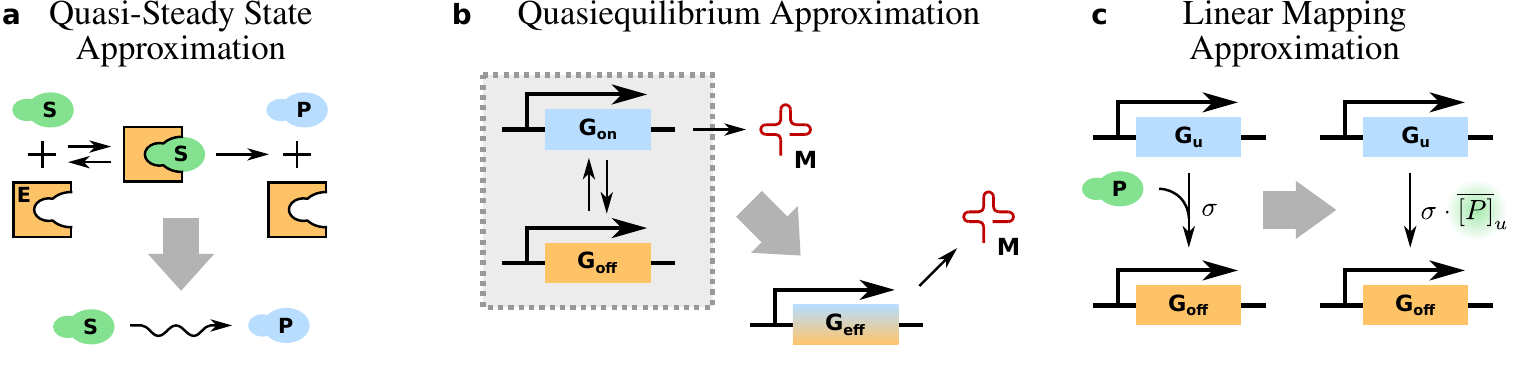}
    \caption{Common model reduction techniques for the Chemical Master Equation. \textbf{(a)} The QSSA eliminates intermediate species which evolve on a faster timescale than others, replacing them by their steady-state values. In the case of the pictured Michaelis-Menten enzyme system this is often applied to the enzyme $E$ and the substrate-enzyme complex $ES$. \textbf{(b)} The QEA is an analogue of the QSSA that can be applied when a reaction and its reverse equilibrate rapidly on timescales of interest. This can occur e.g.~when a gene switches rapidly between states. \textbf{(c)} The LMA replaces protein-gene binding reactions by effective unimolecular reactions, where the concentration of proteins is approximated by its mean conditioned on the gene state. While this does not remove any species from the system, it considerably simplifies the propensities of the reactions.}
    \label{fig:examples}
\end{figure}

The original model $q$ defines a probability distribution over trajectories in $\mathcal X$, and projecting each trajectory onto the chosen reduced state space we get the (exact) projection of $q$ onto this space, which we denote $\tilde q$ (Fig.~\ref{fig:methods}\textbf{a}). This is a stochastic process on $\mathcal {\tilde X}$ that is generally not Markovian and thus cannot be modelled using the CME. We aim to find a tractable approximation $p$ to $\tilde q$ that can be described using the CME, and we will do this by minimising the KL divergence $\KL(\tilde q \, \| \, p)$ between the two models on the space of trajectories. Several well-known examples of model reduction for the CME are illustrated in Fig.~\textbf{\ref{fig:examples}}.

Jumps in $q$ come in two kinds: those that affect the observed species $\tilde \bn$, which we will call visible jumps, and those that only change $\bz$, which we call hidden. The jumps in $\tilde q$ correspond to visible jumps in $q$. In the context of the CME, jumps are typically grouped into reactions with fixed stoichiometry, often also called reaction channels, and we can similarly distinguish visible and hidden reactions in $q$. We will always assume that different reactions have different stoichiometries, so that every jump in $q$ and $p$ corresponds to a unique reaction. Reactions with the same stoichiometry can always be combined by summing their propensities.

We introduce some more notation at this point, which is summarised in Table~\ref{tab:notation} and illustrated in Fig.~\ref{fig:methods}\textbf{c}. A single realisation, or trajectory, of $q$ is defined by the sequence of states $\bn_0, \bn_1, \bn_2, \ldots \in \mathcal X$ visited and jump times $0 < t_1 < t_2 < \ldots$. We will write $\bn_{[0,T]} = \{\bn(t)\}_{0 \leq t \leq T}$ for a trajectory, where $\bn(0) = \bn_0$ and $\bn(t) = \bn_i$ for $t_i \leq t < t_{i+1}$, and denote by $\tau_i = t_{i+1} - t_i$ the waiting times between jumps.

For a continuous-time Markov process $p$, e.g.~one defined using the CME, we denote the transition rate from state $\bn$ to $\bnp \neq \bn$ by $p_{\bnp \leftarrow \bn}$. We let $p_{\leftarrow \bn} = \sum_{\bnp \neq \bn} p_{\bnp \leftarrow \bn}$ be the total transition rate out of $\bn$. The transition probabilities at $\bn$ are then given by $p_{\bnp | \bn} = p_{\bnp \leftarrow \bn } / p_{\leftarrow \bn}$. For completeness we define $p_{\bn|\bn} := 0$. 

\label{sec:notation}
{
\centering
\begin{table}[b]
    \centering
    \begin{tabular}{ c l @{\hspace{1cm}} c l }
    \toprule 
    \textbf{Symbol} & \textbf{Explanation} & \textbf{Symbol} & \textbf{Explanation} \\
    \midrule 
        $q$ & Full model & $q_{\bnp \leftarrow \bn}$ & Transition rate \\
        $\tilde q$ & Projected model & $q_{\leftarrow \bn}$ & Total transition rate \\
        $p$ & Reduced model & $q_{\bnp \given \bn}$ & Transition probability \\
    \midrule
        $\bn$ & Full state vector & $q_0(\bn)$ & Initial distribution  \\
        $\tilde \bn$ & Reduced state vector & $q_t(\bn)$ & Single-time marginal \\
        $\bz$ & Unobserved state vector && \\
    \bottomrule
    \end{tabular}
    \caption{Notation used in this paper.}
    \label{tab:notation}
\end{table}
}

\subsection{Minimising the KL Divergence}

The Kullback-Leibler Divergence between two distributions $\tilde q$ and $p$ is defined as
\begin{align}
    \KL(\tilde q \, \| \, p) &= \int \tilde q(x) \log \frac{\tilde q(x)}{p(x)} \dif x = \E_{\tilde q}\left[\log \tilde q(x)\right] - \E_{\tilde q}\left[\log p(x)\right] \label{eq:kl} \\
    &= \E_{\tilde q}\left[\log \tilde q(x)\right] + H(\tilde q; p)
\end{align}

\noindent where the quantity $H(\tilde q; p)$ is known as the cross-entropy:
\begin{align}
    H(\tilde q; p) &= - \E_{\tilde q}\left[\log p(x)\right] \label{eq:xent}
\end{align}
    
Minimising the KL divergence with respect to $p$ is therefore equivalent to minimising the cross-entropy. Looking at the cross-entropy we see that  minimising the KL divergence is a standard inference problem: maximise the (average) log-likelihood of samples drawn from $\tilde q$. 

In this paper samples from $\tilde q$ and $p$ are trajectories. We therefore need to compute the log-likelihood of an arbitrary trajectory $\bn_{[0,T]}$ under $p$, which by assumption is a stochastic reaction network described by the CME. The log-likelihood of a trajectory can be computed as a product of the transition probabilities as is done in Appendix~\ref{apdx:traj_likelihood}. For a trajectory visiting the states $\bn_0, \bn_1, \ldots, \bn_k$ with waiting times $\tau_0, \tau_1, \ldots, \tau_k$ the log-likelihood is given by
\begin{align}
    \log p(\bn_{[0,T]}) &= \log p_0(\bn_0) - \sum_{i=0}^k \tau_i \, p_{\leftarrow \bn_i} + \sum_{i=1}^k \log p_{\bn_i \leftarrow \bn_{i-1}} \label{eq:traj_likelihood}
\end{align}

\noindent This expression can also be obtained by considering the probability of each transition in the Gillespie algorithm. In the presence of time-dependent rates the equation generalises to
\begin{align}
    \log p(\bn_{[0,T]}) &= \log p_0(\bn_0) - \sum_{i=0}^k \int_{t_i}^{t_{i+1}} \!\! p_{\leftarrow \bn_i}(s) \,\dif s + \sum_{i=1}^k \log p_{\bn_i \leftarrow \bn_{i-1}}(t_i), \label{eq:traj_likelihood_td}
\end{align}

\noindent where $t_1, \ldots, t_k$ are the jump times and we set $t_{k+1} = T$ for convenience.

Note that computing the likelihood of a fully observed trajectory is easier than computing the likelihood of a trajectory sampled at discrete time points, which is the form of data typically observed in biological experiments. The difference is that computing the likelihood in the latter case requires integrating over all possible courses of the trajectory between observations; this integral is computed implicitly by the Chemical Master Equation, which is generally hard to solve. For a fully observed trajectory there are no hidden variables to integrate out, which considerably simplifies the task of computing likelihoods.

Returning to the problem of minimizing the KL divergence \eqref{eq:kl}, or equivalently the cross entropy, we need to compute expectations over $\tilde q$ that are not in general available in closed form. We can instead approximate them by simulating $N$ trajectories $\tilde \bn^{(i)}_{[0,T]}$, $i = 1, \ldots, N$, from $\tilde q$ and computing the estimate
\begin{align}
    \widehat H(\tilde q; p_\theta) &= -\frac 1 N \sum_{i=1}^N \log p(\tilde \bn^{(i)}_{[0,T]}). \label{eq:xent_mc}
\end{align}

\noindent We can then minimise this with respect to the reduced model parameters by gradient descent. Formul\ae{} for the gradients can be computed by differentiating Eqs.~\eqref{eq:traj_likelihood} or \eqref{eq:traj_likelihood_td} by hand, or by using automatic differentiation software. Minimising the estimated cross-entropy via gradient descent yields a probabilistic estimate for the optimal parameters, which can be expected to converge to the true solution as $N \rightarrow \infty$. Summarising the contents of this section we arrive at Algorithm~\ref{alg:klopt} for automatically fitting the optimally reduced model.

\begin{algorithm}[H]
\caption{Model reduction via stochastic gradient descent.}
\begin{algorithmic}
\State \textbf{Inputs:} number of simulations $N$, simulation length $T$, full model $q$, model family $\{p_\theta\}$, learning rate $\eta$
\State \textbf{Output:} reduced model parameters $\widehat \theta$
\algrule
\ForAll{$i = 1, \ldots, N$}
\State \textbf{sample} $\bn^{(i)}_{[0,T]}$ \textbf{from} $q$
\State \textbf{project} $\bn^{(i)}_{[0,T]}$ \textbf{to} $\tilde \bn^{(i)}_{[0,T]}$
\EndFor
\State \textbf{end}
\State \textbf{initialize parameters} $\hat \theta$
\While{not converged}
\State \textbf{compute} $\widehat H(\tilde q; p_{\hat \theta})$ \textbf{using} \eqref{eq:traj_likelihood_td}, \eqref{eq:xent_mc}
\State \textbf{compute} $\nabla_\theta \, \widehat H(\tilde q; p_{\hat \theta})$
\State $\hat \theta \mathrel{-}= \eta \, \nabla_\theta \, \hat H(\tilde q; p_{\hat \theta})$ 
\EndWhile
\Return $\hat \theta$
\end{algorithmic}
\label{alg:klopt}
\end{algorithm}

\subsubsection*{Example: Telegraph Model}

We will illustrate the above by means of a well-known example, the telegraph model of gene transcription \cite{peccoud_markovian_1995}, simplified to neglect degradation and illustrated in Fig.~\ref{fig:tg}. The full model $q$ consists of three species: a gene in the on state ($G_{\mathrm{on}}$) and the off state ($G_{\mathrm{off}}$), together with mRNA ($M$). The gene switches between both states with constant activation rate $\sigma_{\mathrm{on}}$ and inactivation rate $\sigma_{\mathrm{off}}$, and in the active state produces mRNA with rate $\rho_{\mathrm{on}}$. The reduced model $p$ consists only of mRNA ($M$), which is produced at constant rate $\rho_{\mathrm{eff}}$. 

In this example a state in the full model can be represented as $\bn = (g,m)$, where $g$ is the gene state (on or off) and the $m$ is the number of mRNA present, while a reduced state $\tilde \bn = (m)$ is given by the number of mRNA molecules only, with the unobserved component $\bz = (g)$ being the gene state. The only reaction in $q$ that descends to $\tilde q$ is mRNA production, whereas the two gene switching reactions are not observed. Note that the projection $\tilde q$ of the telegraph model is not Markovian, as the instantaneous mRNA production rate depends on the current gene state $g$, but we can sample from $\tilde q$ by simulating $q$ and discarding the information about the gene state.

The reduced model $p$ is a Poisson process whose only parameter is the production rate $\rho_{\mathrm{eff}}$. The only trajectories possible under $p$ are those starting at $m = 0$ and increasing by one at every jump. The log-likelihood of such a trajectory $\tilde \bn_{[0,T]}$ under $p$ is given by
\begin{align}
    \log p(\tilde \bn_{[0,T]}) &=  \tilde \bn(T) \log \rho_{\mathrm{eff}} - \rho_{\mathrm{eff}} T, \label{eq:example_likelihood_p}
\end{align}

\noindent where $\tilde \bn(T)$ is the total number of mRNA produced up to time $T$, as can be verified using Eq.~\eqref{eq:traj_likelihood}. The log-likelihood of any other trajectory is~$-\infty$. 

Note that Eq.~\eqref{eq:example_likelihood_p} describes the likelihood for an entire trajectory, as opposed to the likelihood of observing $\bn_T$ molecules at time $T$, which follows a Poisson distribution with rate $\rho_{\mathrm{eff}} T$ and can be obtained by integrating Eq.~\eqref{eq:traj_likelihood} over all trajectories that end at the same $\bn_T$. Indeed, conditioned on observing $\bn_T$ mRNA molecules at time $T$ their individual production times are uniformly and independently distributed on $[0,T]$ by the properties of the Poisson process, and integrating Eq.~\eqref{eq:example_likelihood_p} over all possible combinations of production times we recover the usual Poisson likelihood, keeping in mind that any permutation of the production times yields the same trajectory.

The reduced model $p$ can model every trajectory obtained from $\tilde q$, so the log-likelihood of any sample from $\tilde q$ is finite. The mRNA transcription reactions in both models correspond. If $q$ were to include e.g.~mRNA degradation as is usual in the literature, some trajectories from the full model would feature decreases in $\tilde \bn$ and be impossible under $p$; in this case the KL divergence would infinite, which signals that the reduced model $p$ is not appropriate and needs to be extended to include degradation. 

For this simple example the likelihood of a reduced trajectory under $p$ only depends on the total mRNA produced. We can therefore compute the cross-entropy explicitly:
\begin{align}
    H(\tilde q; p) &= \E_q\left[ \tilde \bn(T) \right] \, \log\rho_{\mathrm{eff}} - \rho_{\mathrm{eff}} \, T.
\end{align}

\noindent Minimising this with respect to $\rho_{\mathrm{eff}}$ yields the optimum
\begin{align}
    \rho_{\mathrm{eff}}^* &= \frac 1 T \, \E_q\left[ \tilde \bn(T) \right], \label{eq:example_rhoeff}
\end{align}

\noindent which is the average mRNA production rate on the interval $[0,T]$. Thus the optimal approximation to the telegraph model is obtained by setting the mRNA production rate to its mean value. This is the result we obtain by applying the QEA to this system (see \ref{sec:theory_qea}), which suggests that the approximation will be better if $\sigma_{\mathrm{on}}$ and $\sigma_{\mathrm{off}}$ are large compared to $\rho_{\mathrm{on}}$. We will see how to evaluate the approximation quality in \ref{sec:theory_kldiv}.

\subsection{Analysing the KL Divergence}

Having derived a numerical procedure to minimise the KL divergence \eqref{eq:kl}, or equivalently the cross-entropy \eqref{eq:xent}, we analyse these quantities to understand how information-theoretic model reduction works. As shown in Appendix~\ref{apdx:kl_mc}, for a Markovian reduced model $p$ the cross-entropy can be written as
\begin{align}
    H(\tilde q; p)_{[0,T]} &= H(\tilde q_0; p_0) +  \int_0^T \sum_{\tilde \bn} \tilde q_{t}(\tilde \bn) \left( p_{\leftarrow \tilde \bn}(t) - \sum_{\tilde \bnp \neq \tilde \bn} \tilde q_{\tilde \bnp \leftarrow \tilde \bn}(t) \log p_{\tilde \bnp \leftarrow \tilde \bn}(t)\right) \dif t \label{eq:xent_ct} \\
    &= H( \tilde q_0; p_0) + \int_0^T H(\tilde q; p)_t \, \dif t,
\end{align}

\noindent with the instantaneous cross-entropy rate at time $t$ defined as
\begin{align}
    H(\tilde q; p)_t &= -\sum_{\tilde \bn} \tilde q_{t}(\tilde \bn) \left( \sum_{\tilde \bnp \neq \tilde \bn} \tilde q_{\tilde \bnp \leftarrow \tilde \bn}(t) \log p_{\tilde \bnp \leftarrow \tilde \bn}(t) - p_{\leftarrow \tilde \bn}(t) \right).
\end{align}

\noindent Here we use the effective transition rates
\begin{align}
\tilde q_{\tilde \bnp \leftarrow \tilde \bn}(t) = \lim_{\delta t \rightarrow 0} \frac 1 {\delta t} P(\tilde q(t + \delta t) = \tilde \bnp \, | \, \tilde q(t) = \tilde \bn, q_0) \qquad (\tilde \bnp \neq \tilde \bn). \label{eq:def_trans_rate}
\end{align}

\noindent If $\tilde q$ is not Markovian, the transition probability from a state $\tilde \bn$ at time $t$ will be affected by the history of the process, and hence on the initial distribution $q_0$. If $\tilde q$ is Markovian, Eq.~\eqref{eq:def_trans_rate} reduces to the classical transition rate which is independent of $q_0$.

Looking at Eq.~\eqref{eq:xent_ct} we make two important observations. As intimated in the discussion of the telegraph example of the previous section, if any trajectory under $\tilde q$ has jumps which are not allowed under $p$, i.e.~if $\tilde q_{\tilde \bnp \leftarrow \tilde \bn} \neq 0$ while $p_{\tilde \bnp \leftarrow \tilde \bn} = 0$, then the cross-entropy \eqref{eq:xent_ct} will be $\infty$: the reduced model $p$ is not flexible enough to model $\tilde q$. On the other hand, if $p$ contains transitions which are impossible under $\tilde q$, i.e.~if $p_{\tilde \bnp \leftarrow \tilde \bn} \neq 0$ while $\tilde q_{\tilde \bnp \leftarrow \tilde \bn} = 0$, then these can only increase the cross-entropy \eqref{eq:xent_ct}. This means that the optimal reduced model $p$ does not contain transitions beyond those in $\tilde q$, and in the context of the CME we can therefore assume that $p$ and $\tilde q$ have the same reactions, with different propensities (Markovian for $p$, not necessarily for $\tilde q$).

If the $i$-th reaction in $p$ has propensity function $\rho_i(\tilde \bn; t)$ and net stoichiometry $\bm{s}_i$ we obtain the following decomposition of the cross-entropy rate at time $t$:
\begin{align}
    H(\tilde q; p)_t &= -\sum_i \sum_{\tilde \bn} \tilde q_{t}(\tilde \bn) \left[ \tilde q_{\tilde \bn + \tilde{\bm{s}_i} \leftarrow \tilde \bn} \log{\rho_i(\tilde \bn; t)} - \rho_i(\tilde \bn; t) \right], \label{eq:kl_reactions}
\end{align}

\noindent where the first sum is over all reactions in $p$, or equivalently all visible reactions in $q$. The total cross-entropy is obtained by integrating Eq.~\eqref{eq:kl_reactions} over $[0,T]$, and we can find the optimal $p$ by optimising the cross-entropy for each reaction separately.

We can minimize Eq.~\eqref{eq:kl_reactions} analytically if the full model $q$ is Markovian. Assume there is precisely one reaction in $q$ with net stoichiometry $\tilde{\bm{s}_i}$ in the projection, and let $\sigma_i$ be its propensity function. Differentiating the above equation with respect to $\rho_i(\tilde \bn; t)$ and setting the derivative to zero we obtain
\begin{align}
    \rho_i^*(\tilde \bn; t) &= \sum_\bz q_{t}(\bz \, | \, \tilde \bn) \, \sigma_i(\tilde \bn, \bz; t) = \E_{\bz}\!\left[\sigma_i(\tilde \bn, \bz; t) \given \tilde \bn; t\right]. \label{eq:propensity_optimal}
\end{align}

\noindent The optimal propensity for a reaction under $p$ is the expected propensity under the original model conditioned on the observed state $\tilde \bn$. In particular, if the propensity of the original reaction does not depend on unobserved species it can be taken over directly.

In practice we often place constraints on the reduced propensities $\rho_i(\tilde \bn; t)$ such as time-homogeneity and mass-action kinetics, which result in constrained optima. For example, if $\rho_i$ is taken to be independent of time we have to integrate Eq.~\eqref{eq:kl_reactions} over $[0,T]$, and as $T \rightarrow \infty$ it can be verified that the total cross-entropy is minimised for
\begin{align}
    \rho_i^*(\tilde \bn) &= \sum_\bz q_\infty(\bz \, | \, \tilde \bn) \, \sigma_i(\tilde \bn, \bz) = \E_{\bz}\!\left[\sigma_i(\tilde \bn, \bz) \given \tilde \bn; t = \infty \right], \label{eq:propensity_optimal_ss}
\end{align}

\noindent which is the steady-state version of Eq.~\eqref{eq:propensity_optimal}.

For other forms of propensity functions the optimum can generally be obtained by a similar averaging procedure. In all cases the main difficulties lie in estimating the relevant conditional expectations, which can be done numerically using the Monte Carlo approach presented in the previous section. Eq.~\eqref{eq:propensity_optimal} is perhaps the central result of this paper, and we will see that it coincides with the propensities obtained using the QSSA, the QEA and the LMA in when these are applied. These methods can therefore all be seen as special cases of our variational approach based on minimising the KL divergence.

\subsubsection*{Example: Telegraph Model, cont.}

\begin{figure}
    \centering
    \includegraphics{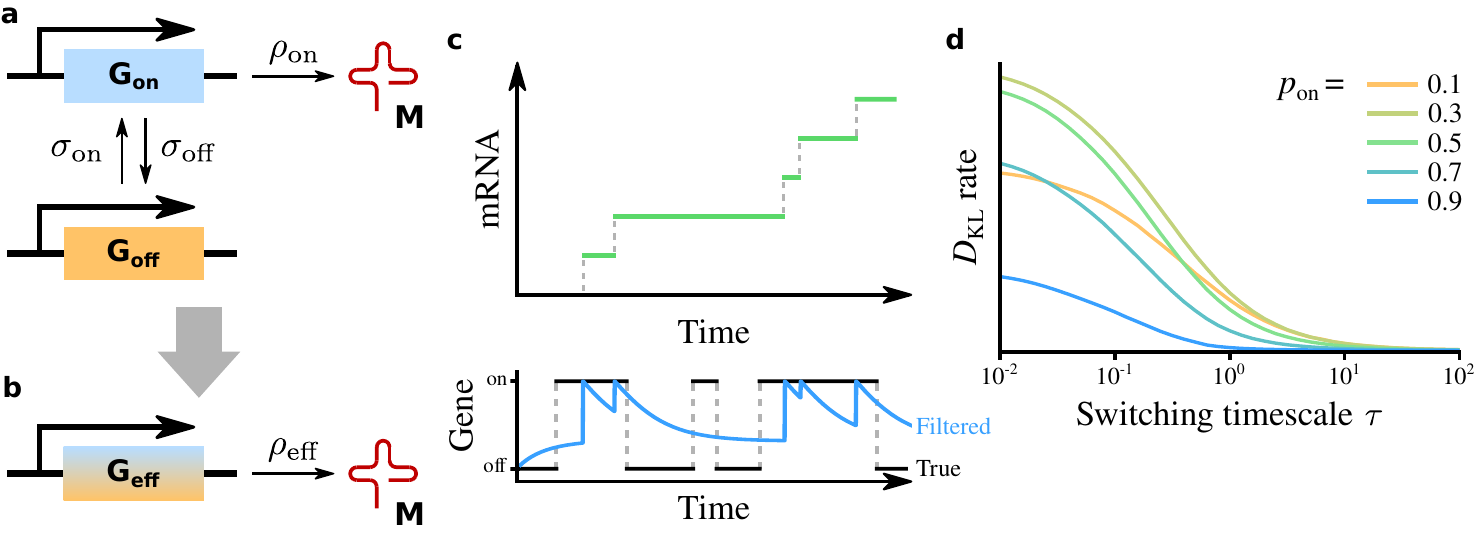}
    \caption{Model reduction illustrated using the telegraph model without degradation. \textbf{(a)} Schematic of the model. \textbf{(b)} The reduced model is equivalent to a Poisson process with rate $\rho_{\mathrm{eff}}$. \textbf{(c)} Example trajectory from the telegraph model, showing mRNA numbers (top) and the gene state (bottom) in time. If we only observe mRNA numbers we can infer the current gene state by filtering, obtaining the probability for the gene being on at a given time (blue line). When mRNA numbers increase the on probability jumps to $1$ since transcription only happens in that state. \textbf{(d)} KL divergence rate between the telegraph model and its Poisson reduction, for various choices of the switching timescale $\tau = \sigma_{\mathrm{on}} + \sigma_{\mathrm{off}}$ and on probability $p_{\mathrm{on}} = \sigma_{\mathrm{on}}/(\sigma_{\mathrm{on}} + \sigma_{\mathrm{off}})$, assuming a fixed transcription rate $\rho_{\mathrm{on}} = 1$. The Poisson approximation becomes more accurate as the switching timescale increases compared to the transcription timescale.}
    \label{fig:tg}
\end{figure}

Consider the telegraph model from the previous section. If we make the reduced model $p$ more flexible by allowing a nonlinear and time-dependent propensity $\rho(m;t) = \rho_{\mathrm{eff}}(m;t)$, then the cross-entropy between $\tilde q$ and $p$ can be computed as
\begin{align}
    H(\tilde q; p) &= \int_0^T \sum_m \tilde q_{t}(m) \, \big(\tilde q_{m+1 \leftarrow m}(t) \log \rho_{\mathrm{eff}}(m; t) - \rho_{\mathrm{eff}}(m; t)\big) \, \dif t. \label{eq:example_xent_td}
\end{align}

\noindent Note that both $\tilde q$ and $p$ have the same initial conditions. The effective propensities of $\tilde q$ can be computed as
\begin{align}
    \tilde q_{m+1 \leftarrow m}(t) = \rho_{\mathrm{on}} \, q_t(g = \textrm{on} \, | \, m), \label{eq:example_effprop}
\end{align}

\noindent which is $\rho_{\mathrm{on}}$ weighed by the probability that the gene is on at time $t$ given that there are $m$ mRNA molecules present. This is the optimal propensity for the reduced model $p$ in accordance with Eq.~\eqref{eq:propensity_optimal}, featuring a nonlinear dependence on $m$, and this can be shown to be the optimal Markovian approximation to $\tilde q$. 

If we require mass-action kinetics for the reduced model, i.e.~if we let $\rho(m; t) = m\, \rho_{\mathrm{eff}}(t)$ for an effective rate constant $\rho_{\mathrm{eff}}(t)$, minimising the cross-entropy \eqref{eq:example_xent_td} yields
\begin{align}
    \rho_{\mathrm{eff}}(t) &= \rho_{\mathrm{on}} \, q_t(g = \textrm{on}).
\end{align}

\noindent The effective transcription rate learned by $p$ is the expected transcription rate of $q$ at time $t$, this time averaged over all $m$.

Finally if we require $\rho(m; t) = m \, \rho_{\mathrm{eff}}$ to be time-independent we obtain by a similar procedure
\begin{align}
    \rho_{\mathrm{eff}} &= \rho_{\mathrm{on}} \cdot \frac 1 T \int_0^T \!q_t(g = \textrm{on}) \, \dif t, \label{eq:tg_rhoeff_th}
\end{align}

\noindent which is just the previous result averaged over $t$. This is the special case considered in the previous section, where $p$ was restricted to be a Poisson process with constant intensity. As $T \rightarrow \infty$, the optimum~\eqref{eq:tg_rhoeff_th} converges to the expected transcription rate of $q$ at steady state. 

\subsection{Computing KL Divergences}

\label{sec:theory_kldiv}

To this point we have been occupied with minimising the KL divergence \eqref{eq:kl}, or rather the cross-entropy \eqref{eq:xent}, with respect to $p$. In order to assess the performance of the reduced model obtained this way we will compute the full KL divergence explicitly, which requires access to the entropy of the projected model $\tilde q$:
\begin{align}
    H(\tilde q)_{[0,T]} &= -\E_{\tilde \bn_{[0,T]}}\!\left[\log q(\tilde \bn_{[0,T]})\right]. \label{eq:ent}
\end{align}

\noindent As for the cross-entropy the integral in Eq.~\eqref{eq:ent} is generally intractable, and we will approximate the entropy by simulating $N$ samples $\tilde \bn^{(i)}_{[0,T]}$ from $\tilde q$ and computing
\begin{align}
    \widehat H(\tilde q) &= -\frac 1 N \sum_{i=1}^N \log \tilde q(\tilde \bn^{(i)}_{[0,T]}). \label{eq:ent_mc}
\end{align}

Here we face another difficulty, for if $\tilde q$ is not Markovian we cannot use Eqs.~\eqref{eq:traj_likelihood} or \eqref{eq:traj_likelihood_td} to compute log-likelihood of a trajectory $\tilde \bn_{[0,T]}$. Since by assumption $q$ is a Markov process, the projection $\tilde q$ is just a partially observed Markov process and
we can use the so-called forward algorithm for Hidden Markov Models \cite{rabiner_tutorial_1989} to compute the log-likelihood of a trajectory.

The forward algorithm computes the joint probability $q(\tilde \bn_{[0,t]}, \bz_t)$ for the observed trajectory up to time $t$ and the current state of the hidden species, $\bz_t$. As shown in Appendix~\ref{apdx:loglikelihood_marg}, the forward algorithm in this case yields the following set of jump ODEs:
\begin{subequations}
\begin{align}
    q(\tilde \bn_0, \bz_0) &= q_0(\tilde \bn_0, \bz_0), \label{eq:fa_1} \\
    \frac{\dif}{\dif t} q(\tilde \bn_{[0,t]}, \bz_t) &= \sum_{\bz'} q_{(\tilde \bn_t, \bz_t)\leftarrow (\tilde \bn_t, \bz')} \, q(\tilde \bn_{[0,t]}, \bz') - \sum_i \sigma_i(\tilde \bn_t, \bz_t) \, q(\tilde \bn_{[0,t]}, \bz_t), \label{eq:fa_2} \\
    \lim_{t \searrow t_k} q(\tilde \bn_{[0,t]}, \bz_t) &= \sigma_{j_k}(\tilde \bn_k, \tilde z_{t_k}) \cdot \lim_{t \nearrow t_k} q(\tilde \bn_{[0,t]}, \bz'), \label{eq:fa_3} 
\end{align}
\end{subequations}

\noindent with jumps at the jump times of the observed trajectory $\tilde \bn_{[0,T]}$. The first sum in Eq.~\eqref{eq:fa_2} represents hidden reactions (those that do not affect $\tilde \bn$), and the second sum is over all visible reactions in $q$, with propensity functions given by the $\sigma_i$. The marginal likelihood of the observed trajectory can then be computed as
\begin{align}
    \tilde q(\tilde \bn_{[0,T]}) &= \sum_{\bz} q(\tilde \bn_{[0,T]}, \bz) \label{eq:fa_4}.
\end{align}

The above equations take the form of a modified CME where the observed variables are fixed. They can be solved using an adaptation of the Finite State Projection algorithm \cite{munsky_finite_2006}, and in cases where the unobserved state space is finite, exact solutions can be computed numerically. We summarise the procedure to estimate the KL divergence defined in Eq.~\eqref{eq:kl} in Algorithm~\ref{alg:klcomp}.
\begin{algorithm}[H]
\caption{Computing the projected KL divergence between a full and a reduced model.}
\begin{algorithmic}
\State \textbf{Inputs:} number of simulations $N$, simulation length $T$, full model $q$, reduced model $p$
\State \textbf{Output:} KL divergence estimate $\widehat \KL(\tilde q \, \| \, p)$
\algrule
\State $\hat L := 0$
\ForAll{$i = 1, \ldots, N$}
\State \textbf{sample} $\bn_{[0,T]}$ \textbf{from} $q$
\State \textbf{project} $\bn_{[0,T]}$ \textbf{to} $\tilde \bn_{[0,T]}$
\State \textbf{compute} $p(\tilde \bn_{[0,T]})$ \textbf{using} \eqref{eq:traj_likelihood_td}
\State \textbf{compute} $\tilde q(\tilde \bn_{[0,T]})$ \textbf{using} \eqref{eq:fa_1}--\eqref{eq:fa_3}, \eqref{eq:fa_4}.
\State $\hat L \mathrel{+}= \tilde q(\tilde \bn_{[0,T]}) - \tilde p(\tilde \bn_{[0,T]})$
\EndFor 
\Return $\hat L / N$
\end{algorithmic}
\label{alg:klcomp}
\end{algorithm}

Rearranging Eqs.~\eqref{eq:fa_1}--\eqref{eq:fa_3} we can show that the marginal distribution itself solves the following jump ODE:
\begin{subequations}
\begin{align}
    \log \tilde q(\tilde \bn_{0}) &= \log \tilde q_0(\tilde \bn(0)), \label{eq:mc_loglikelihood_marg_ode_1} \\
    \frac{\dif}{\dif t} \log \tilde q(\tilde \bn_{[0,t]}) &= -\sum_i \E_{\bz_t}\!\left[\sigma_i(\tilde \bn(t), \bz_t) \given \tilde \bn_{[0,t]}, q_0; t \right], \label{eq:mc_loglikelihood_marg_ode_2} \\
    \lim_{t \searrow t_k} \log \tilde q(\tilde \bn_{[0,t]}) &= \lim_{t \nearrow t_k} \left(\log \tilde q(\tilde \bn_{[0,t]}) + \log \E_{\bz_t}\!\left[\sigma_{j_k}(\tilde \bn_k, \tilde z_t) \given  \tilde \bn_{[0,t]}, q_0; t\right] \right). \label{eq:mc_loglikelihood_marg_ode_3}
\end{align}
\end{subequations}

\noindent Eqs.~\eqref{eq:mc_loglikelihood_marg_ode_1}--\eqref{eq:mc_loglikelihood_marg_ode_3} generalise Eqs.~\eqref{eq:traj_likelihood} and \eqref{eq:traj_likelihood_td} to the presence of unobserved species $\bz$. 

The equations for the log-likelihood are linear and can be split into a sum of contributions for each reaction in $\tilde q$. Integrating over all reduced trajectories $\tilde \bn_{[0,T]}$ shows that the entropy $H(\tilde q)$ can thus be expressed as a sum of contributions from each reaction in $\tilde q$. Eq.~\eqref{eq:kl_reactions} shows that the same decomposition holds for the cross-entropy $H(\tilde q; p)$, so entire KL divergence $\KL(\tilde q \,\| \,p)$ can be decomposed into contributions coming from the individual reactions, which allows us to differentially analyse the accuracy of the reduced model $p$ for each reaction separately. We note that all reactions together determine the distribution $q$ on trajectories over which we integrate, so this decomposition is not strict in practice.

\subsubsection*{Example: Telegraph Model (cont. 2)}

We can solve the filtering problem for the telegraph model exactly as the unobserved state space is $2$-dimensional. Letting $F(i; t) = q\!\left(g(t) = i, m_{[0,t]}\right)$, Eq.~\eqref{eq:fa_2} reduces to
\begin{align}
    \frac{\dif}{\dif t} \, F(\mathrm{on}; t) &= -(\rho_{\mathrm{on}} + \sigma_{\mathrm{off}}) \, F(\mathrm{on}; t) + \sigma_{\mathrm{on}} \, F(\mathrm{off}; t), \\
    \frac{\dif}{\dif t} \, F(\mathrm{off}; t) &= \sigma_{\mathrm{off}} \, F(\mathrm{on}; t) - \sigma_{\mathrm{on}} \, F(\mathrm{off}; t),
\end{align}

\noindent where $m_{[0,t]}$ denotes the observed mRNA trajectory. At each mRNA production event  we update the probabilities according to Eq.~\eqref{eq:fa_3}:
\begin{align}
    \lim_{t \searrow t_k} F(\mathrm{on}; t) &= \rho_{\mathrm{on}} \cdot \lim_{t \nearrow t_k} F(\mathrm{on}; t), \\
    \lim_{t \searrow t_k} F(\mathrm{off}; t) &= 0.
\end{align}

\noindent The marginal likelihood of the observed mRNA trajectory is then given by
\begin{align}
    q\!\left(m_{[0,t]}\right) &= F(\mathrm{on}; t) + F(\mathrm{off}; t).
\end{align}

As a byproduct the above equations yield the conditional distribution over the gene state at each time $t$ given the trajectory prior to that time point. These are the filtered distributions and illustrated in Fig.~\ref{fig:tg}\textbf{c}, but we only need the marginal likelihood for our purposes. 

Using the above equations to compute the log-likelihood for a large number of trajectories sampled from the telegraph model $q$ yields a numerical approximation of the entropy $H(\tilde q)$, and together with Eq.~\eqref{eq:example_likelihood_p} we obtain an estimate of the KL divergence $\KL(\tilde q \, \| \, p)$. For time-homogeneous propensities the KL divergence will be asymptotically proportional to $T$, and we call the proportionality factor the (steady-state) KL divergence rate. In Fig.~\ref{fig:tg}\textbf{d} we show how this KL divergence rate between the telegraph model and its Poisson approximation changes for various choices of $\sigma_{\mathrm{on}}$ and $\sigma_{\mathrm{off}}$. We observe that the KL divergence rate tends to 0 as the switching rates increase; indeed the Poisson approximation can be obtained from the QEA for this example. 

Our results on the telegraph model without degradation apply almost verbatim when degradation is included: adding the reaction $M \singlereac{} \emptyset$ to the full and reduced models does not affect the KL divergences considered. \emph{A priori} adding degradation changes the probability distribution over trajectories, which can now exhibit decreasing mRNA numbers, but the log-likelihood contributed by the degradation reaction is the same between the telegraph model and its reduction, and therefore the difference cancels in Eqs.~\eqref{eq:mc_loglikelihood_marg_ode_1}--\eqref{eq:mc_loglikelihood_marg_ode_3} and Eq.~\eqref{eq:kl_reactions}. Since the log-likelihood contributed by the gene switching and transcription reactions does not depend on mRNA numbers, which do change in the presence of degradation, the total KL divergence is unaffected. In the presence of feedback, degradation would indirectly affect the KL divergence via its effect on mRNA numbers; we refer the interested reader to the recent paper \cite{moor_dynamic_2022} for an analysis of this type of information flow in stochastic biochemical reaction networks.





\subsection{Relationship with Known Approaches}

\subsubsection{The Quasi-Steady State Approximation}

\label{sec:theory_qssa}

In deterministic chemical kinetics the Quasi-Steady State Approximation (QSSA) is a model reduction technique that can be applied when a system can be partitioned into so-called slow species $\bn^s$ and fast species $\bn^f$ such that the fast species $\bn^f$ evolve on a much faster time scale than the slow species. On the timescale on which the slow species evolve, the fast species can therefore be assumed to reach their steady state almost instantaneously. Thus one assumes $\frac{\dif}{\dif t} \bn^f = 0$, which allows for simplification of the remaining equations for the slow species. The QSSA has famously been applied to Michaelis-Menten enzyme kinetics with the enzyme-substrate complex $ES$ as the fast species, resulting in the classical Michaelis-Menten propensity for product formation (see \ref{sec:exp_mm} for more information). 

The QSSA was extended to the stochastic case in \cite{rao_stochastic_2003}. Here one assumes that conditioned on $\bn^s$, the fast species reach steady state nearly instantaneously compared to the timescale of interest. In our formulation the reduced state space consists of the slow species, $\tilde \bn = \bn^s$, and the unobserved species are the fast species, $\bz = \bn^f$. The QSSA then yields the following reduced CME for the approximation $p$ (see (11) in \cite{rao_stochastic_2003}):
\begin{align}
    \frac{\dif}{\dif t} p_t(\bn^s) &= \sum_{i=1}^t \left[ \tilde \rho_i(\bn^s - \bm S^s_i; t) \, p_t(\bn^s - \bm S^s_i) - \tilde \rho_i(\bn^s; t)\, p_t(\bn^s) \right], \label{eq:cme_qssa}
\end{align}

\noindent where the reduced propensities $\tilde \rho_i$ are defined as
\begin{align}
    \tilde \rho_i(\bn^s; t) &= \E_{\bn^f}\!\left[\rho_i(\bn^s, \bn^f) \given \bn^s; t\right]. \label{eq:qssa}
\end{align}

\noindent  This agrees precisely with Eq.~\eqref{eq:propensity_optimal}, which illustrates how the QSSA can be seen as minimising the KL divergence~\eqref{eq:kl_ct}. We can compare \eqref{eq:qssa} with the true propensities of the projection $\tilde q$, which depend on the entire history of the observed trajectory and the initial distribution $q_0$:
\begin{align}
    \rho_i^{\mathrm{exact}}(\bn; t \, | \, q_0, \bn^s_{[0,t]}) &= \E_{\bn^f}\big[\rho_i(\bn^s, \bn^f) \given \bn^s_{[0,t]}, q_0; t\big].
\end{align}

Some intuition for the relationship between timescale separation and our approach can be gained from equations Eqs.~\eqref{eq:mc_loglikelihood_marg_ode_1}--\eqref{eq:mc_loglikelihood_marg_ode_3}, which describe the probability of a reduced trajectory $\tilde \bn_{[0,T]}$ under $\tilde q$. Reaction propensities under $q$ in general depend on the unobserved species, whose distribution is correlated with the history of the current trajectory. If we assume that the timescale of the unobserved species $\bn^f$ is very fast, then this correlation will decay very quickly and the time-dependence of the conditional distributions in Eqs.~\eqref{eq:mc_loglikelihood_marg_ode_1}--\eqref{eq:mc_loglikelihood_marg_ode_3} will be negligible, so we can replace these equations by
\begin{subequations}
\begin{align}
    \log \tilde q(\tilde \bn_{0}) &= \log \tilde q_0(\tilde \bn(0)), \label{eq:mc_loglikelihood_marg_qea_1} \\
    \frac{\dif}{\dif t} \log \tilde q(\tilde \bn_{[0,t]}) &= -\sum_i \E_{\bz_t}\!\left[\sigma_i(\tilde \bn(t), \bz_t) \given \tilde \bn_t \right], \label{eq:mc_loglikelihood_marg_qea_2} \\
    \lim_{t \searrow t_k} \log \tilde q(\tilde \bn_{[0,t]}) &= \lim_{t \nearrow t_k} \left( \log \tilde q(\tilde \bn_{[0,t]}) + \log \E_{\bz_t}\!\left[\sigma_{j_k}(\tilde \bn_k, \tilde z_t) \given \tilde \bn_t \right] \right).\label{eq:mc_loglikelihood_marg_qea_3}
\end{align}
\end{subequations}

\noindent which is another way of showing that $\tilde q$ has propensities given by \eqref{eq:qssa}.

\subsubsection{The Quasiequilibrium Approximation}

\label{sec:theory_qea}
While the QSSA relies on a reaction network being divisible into slow and fast species that evolve on two different timescales, in practice it is more frequently the case that some \emph{reactions} in a system will be fast and that some will be slow. The Quasiequilibrium Approximation (QEA) developed in \cite{cao_slow-scale_2005,haseltine_origins_2005,goutsias_quasiequilibrium_2005} modifies the QSSA to model this scenario by reformulating the system using extents $\bm a = (a_1, \ldots, a_r)$, where the extent $a_i$ is defined the number of times reaction $i$ has occurred. 

The extents themselves form a Markov chain, and the state of a system can be obtained from its extents as $\bn(t) = \bn_0 + \bm{S} \bm a(t)$, where $\bm S$ is the stoichiometry matrix. The CME of this system is
\begin{align}
    \frac{\dif}{\dif t} q_t(\bm a) &= \sum_i \big[ \rho_i(\bm S \bm a - \bm S_i) \, q_t(\bm a - \bm e_i) + \rho_i(\bm S \bm a) \, q_t(\bm a) \big].
\end{align}

\noindent Here the sum is over reactions and $\bm e_i$ is the vector with a $1$ in the $i$-th position and $0$ elsewhere. The QEA assumes that the extent vector can be divided into slow and fast components $\bm a^s$ and $\bm a^f$, respectively, corresponding to slow and fast reactions, and we define our reduced system to consist only of the slow reactions. From here we can proceed analogously to the QSSA discussed above and obtain that the reduced propensities are given by 
\begin{align}
    \tilde \rho_i(\bm a^s; t) &= \E_{\bm a^f}\!\left[ \rho_i(\bm S \bm a) \given \bm a^s; t\right], \label{eq:qea_prop}
\end{align}

\noindent if the fast reactions can be assumed to equilibrate instantaneously.

A subtlety of this argument is the fact that the fast extents $\bm a^f$ can only increase in time and therefore will not admit a steady-state distribution in general. The conditional means, however, will converge in many cases, e.g.~if $\bm a^f$ consists of both directions of a reversible reaction. If this is the case the QEA yields a well-defined reduction, which moreover agrees with \eqref{eq:propensity_optimal} and therefore minimises the KL divergence to the full model on the space of extents. 

\subsubsection{The Linear Mapping Approximation}

\label{sec:theory_lma}

The Linear Mapping Approximation \cite{cao_linear_2018} replaces a bimolecular reaction of the form $G + X  \singlereac{\sigma} G^*$, where $G$ is a binary species representing a gene state, by a reaction $G \singlereac{\overline \sigma} G^*$ with effective propensity $\overline \sigma$. Assuming mass action kinetics, the propensity function of the bimolecular reaction is $\rho(\bn) = g x \sigma $, where $g$ is the state of $G$ and $x$ the abundance of the species $X$, and the propensity of the linearised version is $\tilde \rho(\bn) = \overline \sigma g$ for an appropriate choice of $\overline \sigma$. Taking the derivative with respect to $\overline \sigma$ of the cross-entropy rate \eqref{eq:kl_reactions} at time $t$ yields
\begin{align}
  \frac{\partial H(\tilde q; p)_t}{\partial \, {\overline \sigma}} &= \sum_{\bn} \displaystyle q_{t}(\bn) \left(g - \frac{\sigma g x}{\overline{\sigma}} \right),
\end{align}

\noindent which vanishes if and only if
\begin{align}
  \overline{\sigma} &= \sigma \cdot \E\left[x \, | \, g = 1; t \right]. \label{eq:lma}
\end{align}

\noindent This provides a mathematical justification for the mean-field assumption underlying \cite{cao_linear_2018}. The LMA approximates this conditional expectation by imposing a self-consistency condition on the linearised system, thereby deriving an effective approximation to \eqref{eq:lma}. 

As an aside we can compute the KL divergence rate between $q$ and the optimal reduction $p$ given by \eqref{eq:lma} analytically to obtain
\begin{align}
    \KL(q \, \| \, p)_t &= \sigma \cdot P(g = 1) \cdot \left( \E\left[ x \log{x} \given g = 1 \right] - \E\left[ x \given g = 1 \right] \E\left[ \log{x} \given g = 1 \right] \right) \label{eq:kl_rate_lma}.
\end{align}

\noindent Here the dependence of the expectations on the time $t$ are suppressed. This expression for the discrepancy incurred by the LMA, which resembles the definition of the variance, quantifies the intuition that the LMA should be accurate if fluctuations of $X$ in the unbound gene state are small. 

The above derivation is not entirely rigorous as the linearisation has a different net stoichiometry than the original reaction and KL divergence between $q$ and $p$ is infinite. To remedy this we can either neglect fluctuations in $X$ due to binding in the original model or consider the KL divergence on a reaction-by-reaction basis as discussed in \ref{sec:theory_kldiv}, since the two reactions correspond despite their different net stoichiometries.

\section{Numerical Experiments}

\subsection{Autoregulatory Feedback Loop}

\label{sec:exp_lma}
\begin{figure}[t]
     \centering
     \makebox[\textwidth][c]{\includegraphics{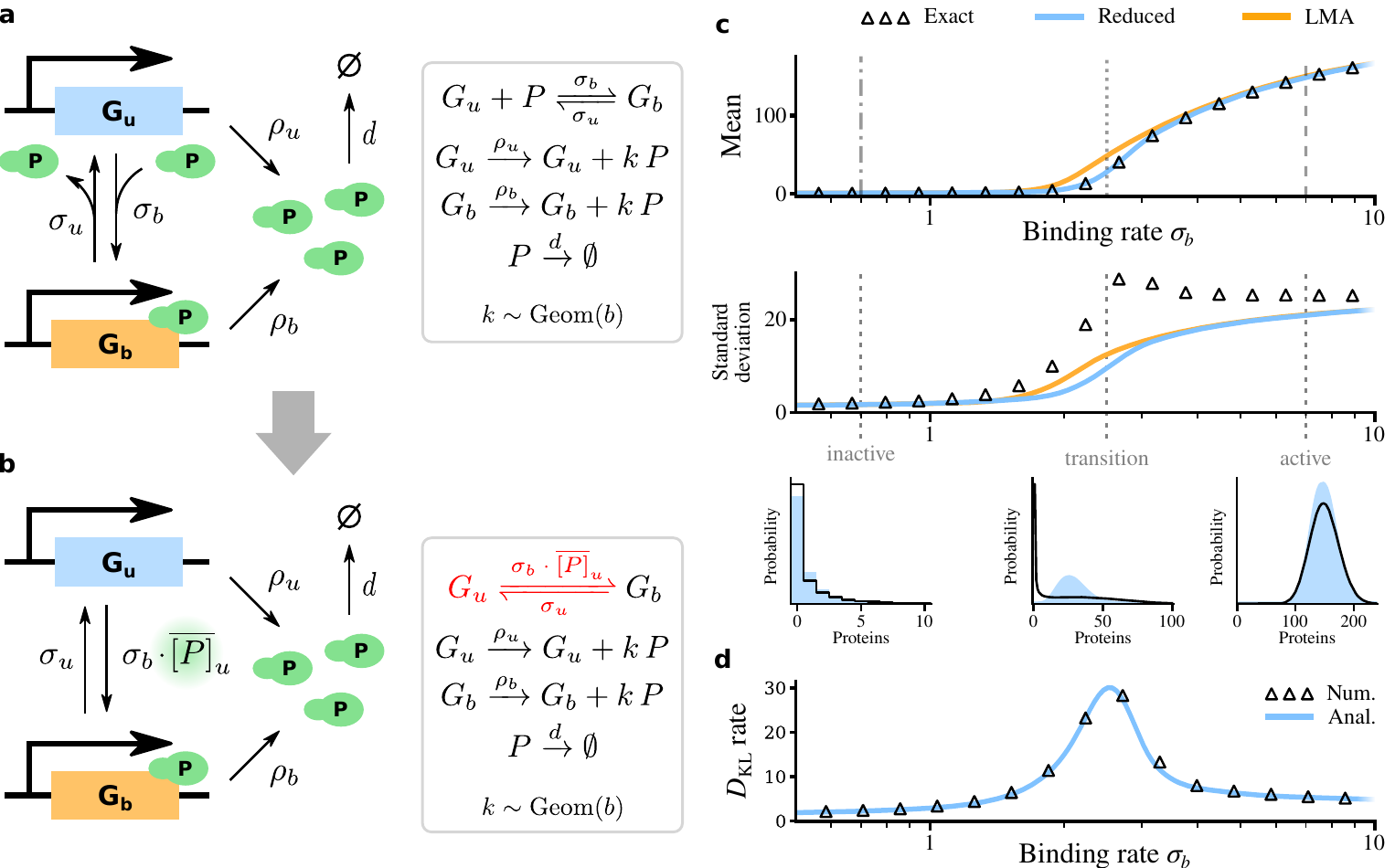}}
     \caption{Reduction of an autoregulatory feedback loop using the Linear Mapping Approximation. \textbf{(a)} Schematic of the full reaction network. The number $k \geq 0$ of proteins produced at each reaction is geometrically distributed with mean $b$. \textbf{(b)} Reduced form of the same reaction system, where the protein-gene binding reaction is replaced by a mean-field approximation. \textbf{(c)} For the chosen parameter values the system exhibits critical behaviour around $\sigma_b = 2.5$, transitioning from a mostly unbound to a mostly bound state. Near the transition protein fluctuations increase and the mean-field assumption in the LMA breaks down. A comparison of steady-state moments and distributions for the full model, the optimal reduction and the LMA shows that the effective reaction rate computed by the latter is generally close to optimal. \textbf{(d)} KL divergence rate at steady state between the full model and the reduced version, computed analytically using Eq.~\eqref{eq:kl_rate_lma} and using Monte Carlo simulations. The peak around the transition matches the observed discrepancy between the full and the reduced model. The remaining model parameters are $\sigma_u = 400$, $\rho_u = 0.3$, $\rho_b = 105$, $d = 1$.}
     \label{fig:lma}
\end{figure}
 
In this section we analyse a simple model of stochastic gene expression featuring positive autoregulation (see Fig.~\ref{fig:lma}\textbf{a}). The system in question consists of a single gene found in two states, bound ($G_b$) and unbound ($G_u$), as well as the coded protein $P$ which can bind to the gene to increase its own transcription rate. For simplicity we do not consider mRNA dynamics explicitly, instead modelling protein production as occurring in geometrically distributed bursts (see \cite{jia_small_2020} for a derivation). The linearised version of the system, using the optimal propensity derived in Section \ref{sec:theory_lma}, is shown in Fig.~\ref{fig:lma}\textbf{b}. In Fig.~\ref{fig:lma}\textbf{c} we compare the steady-state distributions of the full model with its linearisation, where the effective binding rate is computed numerically using our variational approach, and the LMA, where the binding rate is approximated using the self-consistent approach in \cite{cao_linear_2018}. 

This system exhibits critical behaviour for some parameter values (see Fig.~\ref{fig:lma}\textbf{c}), and it was shown in \cite{ocal_parameter_2019} that computing the moments of this system is difficult near points of criticality as most moment closure techniques as well as the LMA yield inaccurate results. Our results show that for all parameters the self-consistent equations of the LMA provides results close to the numerically computed optimum. Yet while both reductions reproduce the mean protein numbers of the full model (the LMA incurring a small bias near the critical point), both fail to capture the increased fluctuations near the critical point, significantly underestimating the variance in protein numbers. Comparing the protein distributions for the full and the optimally reduced model shows a large discrepancy near the critical point, compared to parameters far from it.

We compute the KL divergence rate at steady state between the full model and its linearisation using \eqref{eq:kl_rate_lma} and via Monte Carlo estimation (see Fig.~\ref{fig:lma}\textbf{d}). The steady-state KL divergence rate exhibits a notable peak near the critical point, coinciding with the parameter regime where the linearisation fails to capture the full model. As we move away in either direction from the critical point the KL divergence decreases in accordance with the better approximation of the system by its linearised version. This shows how the KL divergence can be used to assess how well model reduction works for different parameter regimes.

\subsection{Michaelis-Menten Kinetics}

\label{sec:exp_mm}

\begin{figure}
    \centering
    \makebox[\textwidth][c]{\includegraphics{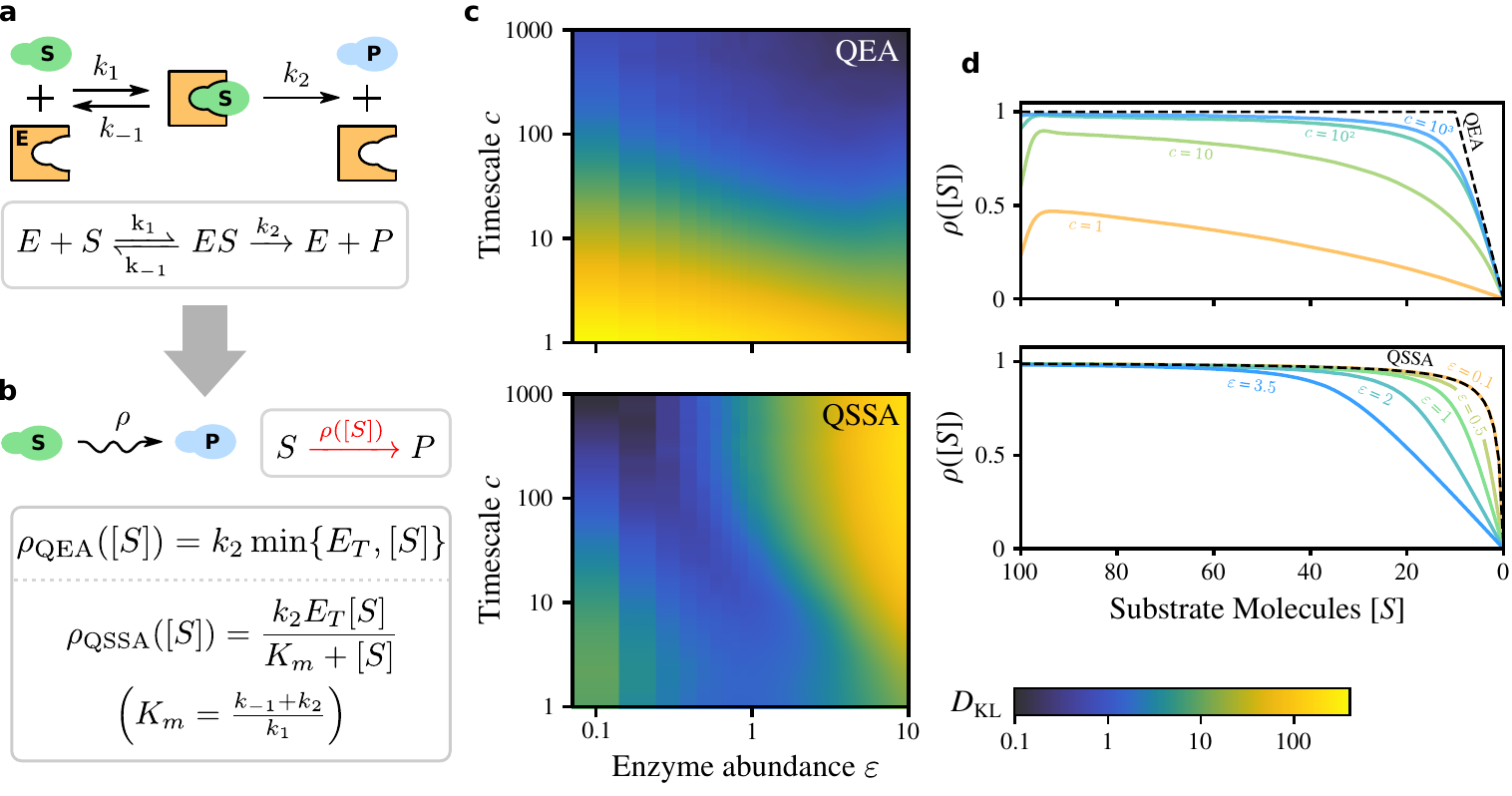}}
    \caption{Comparison of the QEA and the QSSA for Michaelis-Menten kinetics. \textbf{(a)} Schematic of the Michaelis-Menten system. We assume mass action kinetics. \textbf{(b)} Reduction of the Michaelis-Menten system under the QEA and the QSSA yields a one-step process with effective (non-mass action) propensities that depend on the method used. \textbf{(c)} Total KL divergence between the full model and the QEA and QSSA for different values of the timescale parameter $c$ in \eqref{eq:mm_param_c} and the enzyme abundance parameter $\varepsilon$ in \eqref{eq:mm_param_E}. The QEA generally becomes more accurate as $c$ increases, and the QSSA becomes more accurate as $\varepsilon$ decreases. Note the small copy number effects that become apparent in the QSSA for low values of $\varepsilon$ and $c$. \textbf{(d)} Comparison of the effective reaction propensities computed according to Eq.~\eqref{eq:propensity_optimal} with those predicted by the QEA and the QSSA. Here substrates include enzyme-bound substrates. The top figure shows the effective propensities for $\varepsilon = 1$: as $c \rightarrow \infty$ the effective propensities converge to the function predicted by the QEA. The bottom figure shows the effective reaction propensities for $c = 1000$, and as $\varepsilon$ decreases the propensities approach the function predicted by the QSSA. The remaining parameters were fixed to $k_1 = k_{-1} = 0.001$, $k_2 = 0.1$, $E_T = 10$ and $S_T = 100$.} 
    \label{fig:mm}
\end{figure}

We next consider one of the most-studied reaction networks in biology, the Michaelis-Menten model of enzyme kinetics, consisting of an enzyme $E$ and a substrate $S$ that reversibly bind to form an enzyme-substrate complex $ES$ which converts the substrate into the product $P$ (see Fig.~\ref{fig:mm}\textbf{a}). This system is often treated as an example application of both the QSSA and the QEA, which remain the standard model reduction techniques applied to this example. 

The QSSA classically models the enzyme $E$ and the enzyme-substrate complex $ES$ as the fast species, and replaces the full system by a one-step reaction where the substrate is converted to the product with a Hill-type propensity function (see Fig.~\ref{fig:mm}\textbf{b}). This approximation is known to be valid when 
\begin{align}
    E_T \ll [S] + K_m, \label{eq:mm_qssa_accuracy}
\end{align} 

\noindent where $E_T$ is the total number of enzymes and $K_m$ is the Michaelis-Menten constant $(k_{-1} + k_2)/k_1$ \cite{segel_validity_1988,gillespie_legitimacy_2011}. Note that the reduced model is invariant under the scaling
\begin{align}
    E_T &\mapsto \varepsilon \cdot E_T & k_i &\mapsto \varepsilon^{-1} \cdot k_i, \quad i = -1,1,2, \label{eq:mm_param_E}
\end{align}

\noindent where the parameter $\varepsilon$ can be seen as controlling the accuracy of the reduction: the rescaled QSSA condition \eqref{eq:mm_qssa_accuracy} reads
\begin{align}
    \varepsilon E_T \ll [S] + K_m,
\end{align}

\noindent which suggests that small values of $\varepsilon$ should result in the full model more closely resembling the QSSA reduction.

The stochastic QEA for this system is analysed in \cite{goutsias_quasiequilibrium_2005} and also results in a one-step reaction where the propensity function is now piecewise linear (see Fig.~\ref{fig:mm}\textbf{b}). As opposed to the QSSA, the QEA is accurate when substrate binding and unbinding is fast, i.e. $k_1, k_{-1}$ are sufficiently large. Note that the QEA reduction is invariant under the scaling
\begin{align}
    k_1 &\mapsto c \cdot k_1 & k_{-1} &\mapsto c \cdot k_{-1}, \label{eq:mm_param_c}
\end{align}

\noindent where $c$ controls the timescale at which quasiequilibrium is reached, and therefore the accuracy of the QEA. We stress that \eqref{eq:mm_param_E} and \eqref{eq:mm_param_c} are two independent scalings and can be performed simultaneously, as will be the case in Fig.~\ref{fig:mm}\textbf{c}.

In general it may be difficult to predict which of the two approaches is more accurate unless one is clearly in one of the limiting regimes. Based on the KL divergence between the full model and either of the two reactions we can investigate this question for a range of parameters. The reduced model consists of two species, $S$ and $P$, and since $[S] + [P]$ is conserved we can describe it in terms of either of the two. The correct projection for this example identifies the species $S$ and $ES$ in the full model, ie.~we define $[S]_{\mathrm{red}} = [S] + [ES]$ in order to eliminate the binding and unbinding reactions from the projection. This lumping of two rapidly equilibrating species is standard when using the QEA (see e.g.~\cite{jia_dynamical_2020}), but it applies equally well to the QSSA in this case.

In Fig.~\ref{fig:mm}\textbf{b} we compute the total KL divergence over $[0,\infty]$ between the full model and both reductions for a fixed number of substrate molecules. The total KL divergence is finite since all trajectories enter the same absorbing state defined by $[S] = 0$ in a finite amount of time.  As expected for the QEA its accuracy increases with $c$, but whereas the QSSA tends to become less accurate for large $E_T$, in the low enzyme regime we observe a similar decrease in accuracy that is not explained by the deterministic theory.  

The observed decrease is a small copy number effect caused by the fact that the waiting time distribution between two productions of $P$ is not exponential: for very small $E_T$ the unbinding of an enzyme immediately after such an event implies that a significant fraction of such productions occur as a two-step process (where the free enzyme binds another substrate and then converts it) \cite{grima_exact_2017}. For large enough $c$ the binding step is very fast, and assuming that $k_2 \gg k_{-1}$ so that unbinding is unlikely to occur again, the conversion of substrates to products can be viewed as an effective one-step process. The effect of nonexponential waiting time distributions has previously been analysed in \cite{gillespie_subtle_2009}; we see that the KL divergence on trajectories can be sensitive to subtle dynamical effects such as waiting time distribution that are not visible when considering e.g.~the moments of a system, which are well predicted by the QSSA for small values of $c$ and $\varepsilon$. 

Overall we can see that for the chosen parameter values the QSSA generally performs better than the QEA in the regime of small $\varepsilon$, corresponding to low enzyme numbers, and the QEA is most accurate for $c$, keeping in mind the scalings in \eqref{eq:mm_param_E} and \eqref{eq:mm_param_c}. Neither approximation is satisfactory if the number of total enzymes is larger than the amount of substrates, but the binding and unbinding rates are small. In this scenario other approximations, such as the total QSSA \cite{macnamara_stochastic_2008}, which we do not investigate here, will generally be more accurate.

Figures \ref{fig:mm}\textbf{c} and \ref{fig:mm}\textbf{d} compare the effective propensities for the full system, as a function of the unconverted substrate abundance, with the predictions made by the QEA resp.~the QSSA. In the case of the QEA we see that the effective propensities slowly converge to their limit as the timescale parameter $c$ increases. In contrast the QSSA provides a good approximation to the effective propensity as long as the number of substrate molecules is larger than the number of enzymes. While the effective propensities are the optimal choice for the reduced model, the actual quality of the approximation is affected both by the size of the fluctuations of the actual propensities around their mean as well as the degree to which the waiting times of the original system follow an exponential distribution.

\subsection{Genetic Oscillator}

 \begin{figure}
     \centering
     \makebox[\textwidth][c]{\includegraphics{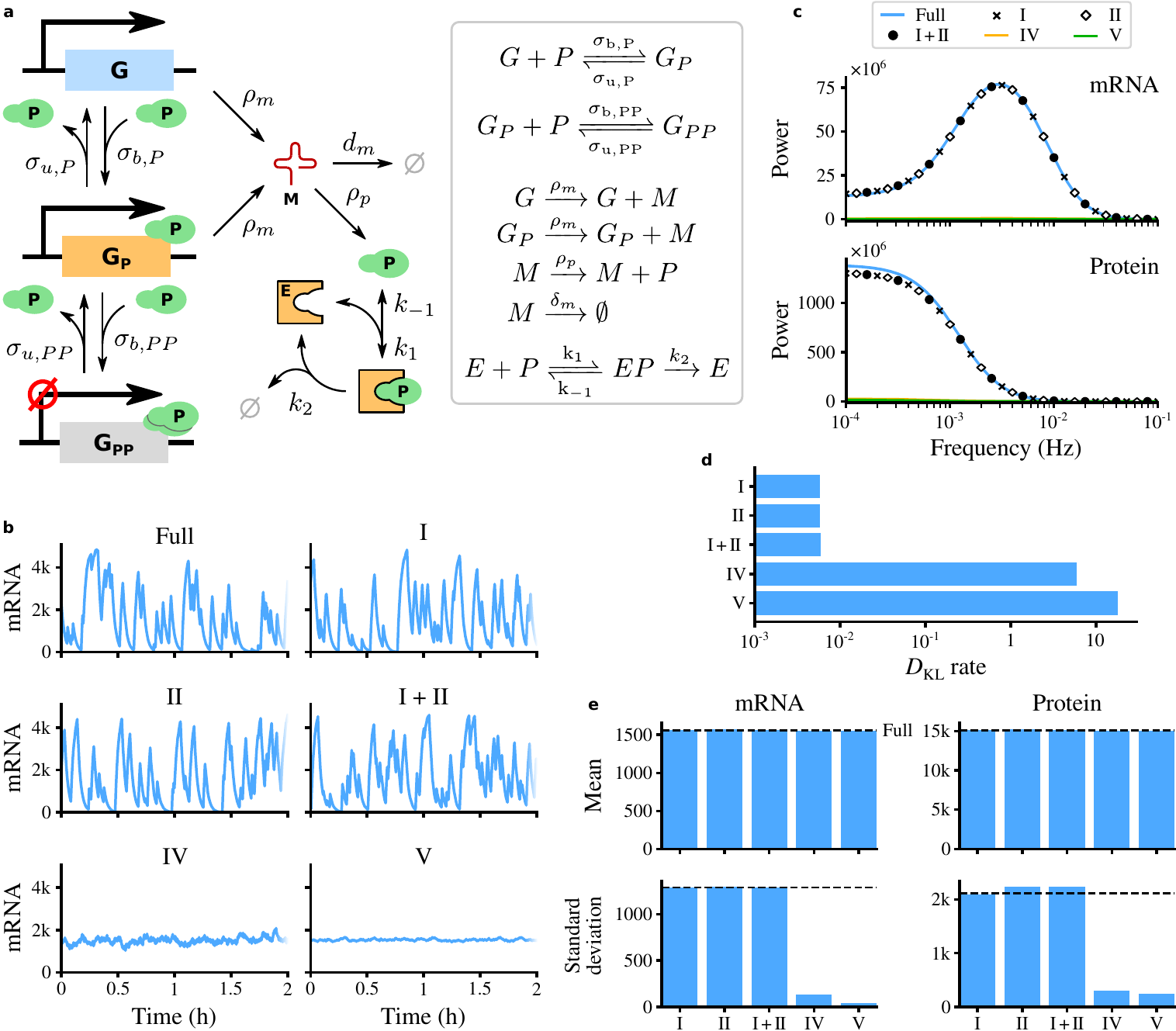}}
     \caption{Comparison of different reductions for the oscillatory gene network. \textbf{(a)} Schematic of the reaction system. We assume mass action kinetics for all reactions in the full model. \textbf{(b)} Example trajectories for the full model and its reductions. Note the oscillatory behaviour in the original model that is not preserved in reductions IV and V. \textbf{(c)} Power spectra of mRNA and protein concentrations (defined as copy numbers normalised by the system size $\Omega$). Models I, II and I+II closely reproduce the oscillatory behaviour of the full model while IV and V do not show sustained oscillations. \textbf{(d)} KL divergence rates between the full model and all reductions. \textbf{(e)} Steady-state means and standard deviations for mRNA and protein numbers. While all reductions closely approximate the mean, Models IV and V do not match the variance of the full model. The parameters for the full model, taken from \cite{thomas_slow-scale_2012}, are $\Omega = 1000$, $E_T = 10$, $\rho_m = 50$, $\rho_p = 0.0045$, $d_m = 0.01$, $k_1 = 0.1$, $k_{-1} = 10$, $k_2 = 10$, $\sigma_{b,P} = 0.001$, $\sigma_{u,P} = 100$, $\sigma_{b,PP} = 1000$, $\sigma_{u,PP} = 1$.}
     \label{fig:osc}
 \end{figure}
 
Our final example is the gene expression system shown in Fig.~\ref{fig:osc}\textbf{a}. This system consists of a gene, mRNA and protein, as well as a Michaelis-Menten type protein degradation mechanism. Up to two protein molecules can bind the gene, and in the twice bound state the gene pauses transcription. This model has been considered in \cite{thomas_slow-scale_2012} as an example system where naive reduction of the CME using Hill-type effective propensities is unable to accurately capture the noise in the original. The true system can exhibit oscillations in mRNA numbers that are caused by the two-step negative feedback and are not present in the reduced version. In this section we want to analyse what reductions can be performed on the full model while still keeping oscillatory behaviour.

We consider five different reductions for this model which are listed in Table~\ref{tab:osc}. Each reduction removes or combines several species such as gene states, thereby reducing the dimensionality of the system. The rightmost column describes the correct projection for each model, derived using an analogous argument as in the Michaelis-Menten example. We fit the unknown (effective) parameters for each model numerically by minimising the KL divergence from the full model. Typical mRNA trajectories for all models can be seen in Fig.~\ref{fig:osc}\textbf{b}. The oscillations in the full model are clearly visible, and they are inherited by reductions I, II and I+II. In contrast, simplifying the first binding step as in IV and V results in a non-oscillatory system (see Fig.~\ref{fig:osc}\textbf{c}). 

\newcommand{\ppp}{{\scriptstyle {P/PP}}}
\newcommand{\newspec}[1]{\textcolor{myblue}{#1}}
\newcommand{\oldspec}[1]{\textcolor{gray}{#1}}
\newcommand{\newrate}[1]{\textcolor{red}{#1}}


In Fig.~\ref{fig:osc}\textbf{d} we compare the KL divergence rates between the full model and reductions I-V (note that the different reductions are defined on different state spaces). While models I, II and I+II have comparatively low KL divergences from the original there is a sharp increase with models IV and V. From this alone one could expect that the latter perform significantly worse, which is indeed the case as shown in Fig.~\ref{fig:osc}\textbf{b}. Fig.~\ref{fig:osc}\textbf{e} presents a comparison of the mean and standard deviations for mRNA and protein abundances; while all models approximate the means very closely, the predicted standard deviations for model IV and V are very far from their true values consistent with the lack oscillatory behaviour. This resembles the results in Section \ref{sec:exp_lma} which also show excellent agreement of the full and reduced model on the mean level, independent of the total approximation quality.

This comparison of various reductions shows how variational model reduction can be employed on a more sophisticated scale, where multiple reductions are possible. Given a list of possible simplifications we can automatically find optimal parameters for each reduction and compute the KL divergence from the true model. Reductions which do not significantly affect the output will generally lead to very low KL divergences compared to those which do, as in this example where a large gap between KL divergence rates separates reductions I, II and I+II from IV and V. The latter do not retain much of the moments or oscillatory dynamics of the full model compared to the other reductions.



\section{Discussion}

In this work we presented an information-theoretic approach to model reduction for the CME based on minimising the KL divergence from the full model and the proposed reduction. Based on this variational principle we derive Eq.~\eqref{eq:propensity_optimal}, showing that the optimal (effective) propensity for a reduction of the Chemical Master Equation equals the mean propensity of the original system conditioned on the state of the reduced system. As a consequence we show that some of the most common approaches to model reduction for the CME - the QSSA \cite{rao_stochastic_2003,kim_relationship_2015,kim_validity_2014}, the QEA \cite{haseltine_approximate_2002,goutsias_quasiequilibrium_2005} and the LMA \cite{cao_linear_2018} - generally try to minimise the KL divergence considered in this paper, providing a general justification for the mean-field arguments proposed in the literature on these methods and connecting them with information theory and likelihood-based approaches. We provide a numerical algorithm for automated fitting of a reduced model based on minimising the KL divergence via stochastic gradient descent, and a numerical algorithm for estimating this KL divergence in order to assess model fit. While the KL divergence between Markov chains has been considered before in e.g. \cite{geiger_optimal_2015,rached_kullback-leibler_2004,amjad_generalized_2020,deng_optimal_2011,opper_variational_2007,wildner_moment-based_2019,bronstein_marginal_2018,moor_dynamic_2022}, to our knowledge it has not been studied in connection with standard model reduction methods as done in this paper. Our numerical results show how the KL divergence can be computed in practice and used in the context of model reduction. 

\begin{landscape}
{
\renewcommand{\singlereac}[1]{\xrightarrow{\hspace{2ex}#1\hspace{2ex}}\rule[-0.9ex]{0pt}{0pt}}
\renewcommand{\doublereac}[2]{\ce{<=>[{#1}][{#2}]}}

\setlength{\tabcolsep}{20pt}
\begin{table}[h]
\centering
{\begin{tabular}{>{\bfseries}c c c r}
    \toprule
    Model & \multicolumn{1}{c}{\textbf{Full reactions}} & \multicolumn{1}{c}{\textbf{Reduced reactions}} & \multicolumn{1}{c}{\textbf{Projection map}} \\
    \midrule
    I & $\oldspec{E} + P \doublereac{}{} \oldspec{EP} \singlereac{} \oldspec{E}$ & $P \singlereac{\textstyle \newrate{d_\mathrm{eff}}} \emptyset$ & \begin{tabular}{r @{ $:=$ }p{3cm}} $[P]_{\mathrm{red}}$ & $[P] + [\oldspec{EP}]$ \end{tabular} \\
    \midrule \\[-1.5ex] 
    II & \makecell{$G + P \doublereac{}{} \oldspec{G_P}$ \\ $\oldspec{G_P} + P \doublereac{}{} \oldspec{G_{PP}}$ \\ $\oldspec{G_P} \singlereac{} \oldspec{G_P} + P$} & \makecell{$G + P \doublereac{\textstyle \sigma_{b,P}}{\textstyle \newrate{\sigma_u}/(\newrate{K_u}+[P])} \newspec{G_{\ppp}}$ \\ \\  $\newspec{G_{\ppp}} \singlereac{\textstyle \newrate{\rho_{\mathrm{eff}}}} \newspec{G_{\ppp}} + P$} & {\renewcommand{\arraystretch}{1.2} \begin{tabular}{r @{ $:=$ }p{3cm}}
        $[P]_{\mathrm{red}}$ & $[P] + [\oldspec{G_{PP}}]$ \\ $[\newspec{G_{\ppp}}]_{\mathrm{red}}$ & $[\oldspec{G_P}] + [\oldspec{G_{PP}}]$
        \end{tabular}}\\
    \\[-1.5ex] \midrule \\[-1.75ex] 
    I + II & (I \& II above) & (I \& II above) & {\renewcommand{\arraystretch}{1.2} \begin{tabular}{r @{ $:=$ }p{3cm}}
        $[P]_{\mathrm{red}}$ & $[P] + [\oldspec{EP}] + [\oldspec{G_{PP}}]$ \\
        $[\newspec{G_{\ppp}}]_{\mathrm{red}}$ & $[\oldspec{G_P}] + [\oldspec{G_{PP}}]$
        \end{tabular}} \\
    \\[-1.75ex] \midrule \\[-1.5ex] 
    IV & \makecell{$\oldspec{G} + P \doublereac{}{} \oldspec{G_P}$ \\[0.1ex] $\oldspec{G_P} + P \doublereac{}{} G_{PP}$ \\ $\oldspec{G} \singlereac{} \oldspec{G} + P$\\ $\oldspec{G_P} \singlereac{} \oldspec{G_P} + P$} & \makecell{$\newspec{G_{-/P}} + P \doublereac{\textstyle \newrate{\sigma_b}[P]/(\newrate{K_b}+[P])}{\textstyle \sigma_{u,PP}} G_{PP}$ \\[1em] $\newspec{G_{-/P}} \singlereac{\textstyle \rho_m} \newspec{G_{-/P}} + P$} & {\renewcommand{\arraystretch}{1.2}\begin{tabular}{r @{ $:=$ }p{3cm}}
         $[P]_{\mathrm{red}}$ & $[P] + [\oldspec{G_P}] + [\oldspec{G_{PP}}]$ \\ $[\newspec{G_{-/P}}]_{\mathrm{red}}$ & $[\oldspec{G}] + [\oldspec{G_P}]$
     \end{tabular}} \\ 
    \\[-1.5ex] \midrule \\[-1.5ex] 
    V & \makecell{$\displaystyle \oldspec{G} + P \doublereac{}{} \oldspec{G_P}$ \\ $\displaystyle \oldspec{G_P} + P \doublereac{}{} \oldspec{G_{PP}}$ \\ $\displaystyle \oldspec{G} \singlereac{} \oldspec{G} + P$\\ $\displaystyle \oldspec{G_P} \singlereac{} \oldspec{G_P} + P$} & \makecell{$\emptyset \singlereac{\displaystyle \frac{\newrate{\alpha} + \newrate{\beta} [P]}{1 + \newrate{\gamma}[P] + \newrate{\delta}[P]^2}} P$} & \begin{tabular}{r @{ $:=$ }p{3cm}} $[P]_{\mathrm{red}}$ & $[P] + [\oldspec{G_P}] + 2[\oldspec{G_{PP}}]$ 
    \end{tabular} \\
    \\[-1.5ex] \bottomrule 
\end{tabular}}
\caption{Five possible reductions for the oscillator model. Model I+II is the combination of the reductions in Models I and II. All propensities are shown without their mass action terms. Species that are lumped or removed in the reduced model are shown in \textcolor{gray}{gray}, and species that do not exist in the full model in \textcolor{myblue}{blue}. Parameters that are introduced in the reduced model are shown in \textcolor{red}{red} and found by minimisation of the KL divergence between the models; all remaining parameters are taken over from the full model. The functional forms of the propensity functions for Models II, IV and V were derived using the QEA.}
\label{tab:osc}
\end{table}
}
\end{landscape}

Using three biologically relevant examples we illustrated how these model reduction techniques can be analysed from this variational perspective. For the autoregulatory feedback loop we showed that the KL divergence provides a useful metric of approximation quality and can detect parameter regimes where the mean-field approximation behind the LMA fails. Using the Michaelis-Menten system we demonstrated how our approach can be used to decide between possible reductions of a model, particularly in the non-asymptotic regime where neither the QSSA nor the QEA are strictly valid. Finally we used the genetic oscillator in \cite{thomas_slow-scale_2012} as an example to show how different reductions of a given model can be fit automatically, separately and in combination, and assesses which steps in a putative model reduction procedure would impact approximation quality more than others.


One important aspect of model reduction we have not addressed is that of choosing an appropriate architecture for a reduced model. As we investigated in the case of the genetic oscillator, while we can always optimise the parameters given an architecture, the quality of the approximation can greatly depend on which reductions are performed. An algorithm that performs automated model reduction for the CME, computing the maximally reduced version of a given model that still approximates the original within a given threshold, is still missing from the literature. In combination with other machine learning approaches to model reduction such as \cite{caravagna_matching_2016}, however, we hope that our approach may be one of several steps towards such a procedure and. 

While the KL divergence is a well-studied quantity in statistics, it lacks some desirable properties in the context of model reduction. There is in general no simple relationship connecting the KL divergence between two multivariate distributions with those of their marginals, and as a consequence we cannot compute e.g.~the KL divergence between single-time marginals of the full and reduced models from our knowledge of the KL divergence on the trajectory level. Similarly, rigorously establishing a relationship between the KL divergence and dynamical information such as power spectra is difficult. Despite this, our empirical results suggest that the KL divergence on the space of trajectories is a useful quantity that relates to both the discrepancy between single-time marginals and power spectra and can be a useful metric of model fit despite a lack of sharp results. We are optimistic that further investigations in this area will demarcate more clearly which properties of a model are well captured by the KL divergence and which are not.

\section*{Code availability}{Code implementing the methods introduced in this paper can be found at \url{https://github.com/kaandocal/modred}.}

\section*{Acknowledgments}{This work was supported by the EPSRC Centre for Doctoral Training in Data Science (EPSRC grant EP/L016427/1) and the University of Edinburgh for K.~\"O. and a Leverhulme Trust grant (Grant No.~RPG-2020-327) for R.~G.}

\printbibliography

\clearpage 

\appendix

\section{Likelihoods of Fully Observed Trajectories}

\label{apdx:traj_likelihood}
Let $p$ be a continuous-time Markov chain defined on the space $\mathcal X$ with initial distribution $p_0$. If $p$ has time-independent transition rates the Stochastic Simulation Algorithm \cite{gillespie_general_1976}, reproduced in Algorithm~\ref{alg:gillespie}, returns exact samples from $p$. From the description we deduce that the log probability of sampling a given trajectory $\bn_{[0,T]}$ equals
\begin{align}
    \log p(\bn_{[0,T]}) = \log p_0(\bn(0)) - \int_{0}^{T} p_{\leftarrow \bn(s)} \, \dif s + \sum_{i=0}^{k-1} \log(p_{\bn_{i+1}\leftarrow \bn_i}) \label{eq:traj_logl_th}
\end{align}

\noindent where $\bn_0, \bn_1, \ldots$ is the sequence of states visited. If the rates of $p$ are time-dependent one can verify that the log probability is given by the following generalisation of \eqref{eq:traj_logl_th}:
\begin{align}
    \log p(\bn_{[0,T]}) = \log p_0(\bn(0)) - \int_{0}^{T} p_{\leftarrow \bn(s)}(s) \, \dif s + \sum_{i=0}^{k-1} \log(p_{\bn_{i+1}\leftarrow \bn_i}(t_{i+1})),
\end{align}

\noindent were $t_1, t_2, \ldots$ are the jump times.

\begin{algorithm}[H]
\caption{The Stochastic Simulation Algorithm to simulate samples from a Markov chain $p$.}

\begin{algorithmic}
\vspace{5pt}

\State \textbf{Input:} simulation length $T$, Markov chain $p$, initial distribution $p_0$
\State \textbf{Output:} $\bn_{[0,T]}$ - sampled trajectory
\algrule
\State \textbf{sample} $\bn_0 \sim p_0$
\State $t \leftarrow 0$, $i \leftarrow 0$ 
\While{$t < T$}
\State \textbf{sample} $\tau_{i+1} \sim \mathrm{Exp}(1 / p_{\leftarrow \bn_i})$
\State \textbf{sample} $\bn_{i+1} = \bnp$ \textbf{with probability} ${p_{\bnp \leftarrow \bn_i}}/{p_{\leftarrow \bn_i}}$
\State $t \mathrel{+}= \tau_{i+1}$
\EndWhile \\
\Return states $(\bn_0, \bn_1, \ldots)$, jump times $(t_1, t_2, \ldots)$
\end{algorithmic}
\label{alg:gillespie}
\end{algorithm}

\section{Kullback-Leibler Divergence between Markov Chains}

\label{apdx:kl_mc}
In \cite{opper_variational_2007} the authors derive an expression for the Kullback-Leibler divergence between two continuous-time Markov chains $q$ and $p$ defined on a common state space $\mathcal X$. For any integer $N > 1$ let $q^{(N)}$ resp.~$p^{(N)}$ be the time discretisations of $q$ and $p$ with time step $\delta t = T / N$. These define probability distributions on ${\mathcal X}^{N+1}$ with KL divergence
\begin{align}
    \KL(q^{(N)} \, \| \, p^{(N)}) &= \sum_{\bn} q_0(\bn) \log \frac{q_0(\bn)}{p_0(\bn)} + \sum_{i=1}^{N} \sum_{\bn,\bnp} q^{(N)}_{i-1}(\bn) \, q_i^{(N)}(\bnp \, | \, \bn) \log \frac{q_i^{(N)}(\bnp \given \bn)}{p_i^{(N)}(\bnp \given \bn)} \label{eq:dt_kl}
\end{align}

\noindent The transition rates are related to the discrete-time transition probabilities as follows:
\begin{align}
    q^{(N)}_i(\bn \given \bn) &= 1 - (\delta t) \, q_{\leftarrow \bn}(iT/N) + o(\delta t) \\
    q^{(N)}_i(\bnp \given \bn) &= (\delta t) \, q_{\bnp \leftarrow \bn}(iT/N) + o(\delta t) &\qquad  (\bnp \neq \bn) \label{eq:mc_ct_rates}
\end{align}

\noindent Using these identities it can be verified that the KL divergence \eqref{eq:dt_kl} converges to the following as $N \rightarrow \infty$:
\begin{align}
    \KL(q \, \| \, p) &= \KL(q_0 \, \| \, p_0) - \displaystyle \int_0^T \dif t \sum_{\bn} q_{t}(\bn) \, (q_{\leftarrow \bn}(t) - p_{\leftarrow \bn}(t)) \\
    &\qquad + \int_0^T \dif t \sum_{\bnp \neq \bn} q_{t}(\bn)\, q_{\bnp \leftarrow \bn}(t) (\log q_{\bnp \leftarrow \bn}(t) - \log p_{\bnp \leftarrow \bn}(t)) \label{eq:kl_ct}
\end{align}

In general the KL divergence can be written as the difference of the the cross-entropy $H(q; p)$ and the entropy $H(q; p)$, and we define the cross-entropy of two continuous-time Markov chains as 
\begin{equation}
    H(q; p) = H(q_0; p_0) + \displaystyle \int_0^T \dif t \sum_{\bn} q_{t}(\bn) \, p_{\leftarrow \bn}(t) - \int_0^T \dif t \sum_{\bnp \neq \bn} q_{t}(\bn)\, q_{\bnp \leftarrow \bn}(t) \log p_{\bnp \leftarrow \bn}(t).  \label{eq:crossent_ct}
\end{equation}

\noindent The above derivation is valid only if $q_{\bnp \leftarrow \bn} \neq 0$ implies $p_{\bnp \leftarrow \bn} \neq 0$; otherwise the cross-entropy and KL divergence are both infinite.

\section{Likelihoods of Reduced Trajectories}

\label{apdx:loglikelihood_marg}

Let $q$ be a continuous-time Markov chain defined on the discrete state space $\mathcal X$ with initial distribution $q_0$, and let $\tilde q$ be its projection onto $\mathcal {\tilde X}$. Computing the log probability of a trajectory $\tilde \bn_{[0,T]}$ under $\tilde q$ requires integrating over all possible full trajectories $\bn_{[0,T]} = (\tilde \bn_{[0,T]}, \bz_{[0,T]})$ that are compatible with $\tilde \bn_{[0,T]}$, that is,
\begin{align}
    \tilde q(\tilde \bn_{[0,T]}) = \int_{\bz_{[0,T]}} q(\tilde \bn_{[0,T]}, \bz_{[0,T]}) \dif \bz_{[0,T]}
\end{align}

\noindent In order to compute the integral on the right-hand side we consider the time-discretisations of $q$ and $\tilde q$. The marginal likelihood in this case can be obtained using the well-known forward algorithm for HMMs, which sequentially computes the joint probabilities $q^{(N)}(\tilde \bn_{0:i}, \bz_i)$:
\begin{align}
    q^{(N)}(\tilde \bn_0, \bz_0) &= q_0(\tilde \bn_0, \bz_0) \\ 
    q^{(N)}(\tilde \bn_{0:i+1}, \bz_{i+1}) &= \sum_{\bz} q^{(N)}(\tilde \bn_{0:i}, \bz_i = \bz)\,  q^{(N)}(\tilde \bn_{i+1}, \bz_{i+1} \given \tilde \bn_i, \bz_i = \bz)
\end{align}

\noindent The marginal probability of the entire trajectory $\tilde \bn_{0:N}$ can then be computed by marginalising after the last step:
\begin{align}
    q^{(N)}(\tilde \bn_{0:N}) &= \sum_{\bz_N} q^{(N)}(\tilde \bn_{0:N}, \bz_N).
\end{align}

\noindent Passing to the limit $N \rightarrow \infty$ and rewriting the visible transition rates in terms of reaction propensities we arrive at \eqref{eq:fa_1}--\eqref{eq:fa_3}.

\end{document}


\appendix

\section{Likelihoods of Trajectories}

Let $p$ be a continuous-time Markov chain defined on the state space $\mathcal Y$ with initial distribution $p_0$. The Gillespie algorithm\cite{gillespie}, which draws exact samples from $p$, is reproduced as Algorithm \ref{alg:gillespie}. From the description we deduce that the log probability of sampling the returned trajectory $y_{[0,T]}$ equals
\begin{equation}
    \log p(y_{[0,T]}) = \log p_0(y(0)) - \int_{0}^{T} p_{\leftarrow y(s)} \, \dif s + \sum_{i=0}^{k-1} \log(p_{y_{i+1}\leftarrow y_i}) 
\end{equation}

\noindent where $0 = t_0 < t_1 < \cdots < t_k = T$ are the jump times and $y_0, y_1, \ldots$ the states visited. 

This can be extended to the case of a time-inhomogeneous Markov process where the transition rates $p_{y' \leftarrow y} := p_{y' \leftarrow y}(t)$ depend on $t$:
\begin{equation}
    \log p(y_{[0,T]}) = \log p_0(y(0)) - \int_{0}^{T} p_{\leftarrow y(s)}(s) \, \dif s + \sum_{i=0}^{k-1} \log p_{y_{i+1}\leftarrow y_i}(t_{i+1}) \label{eq:logprob_ct_inh}
\end{equation}

For later convenience we analyse the time evolution of the quantity $\log p(y_{[0,t]})$ as a function of $t$. Between jumps it evolves according to the following ODE:
\begin{eqnarray}
    \frac{\dif}{\dif t} \log p(y_{[0,t]}) &= -p_{\leftarrow y(t)}(t) \label{eq:logprob_ev_cont}
\end{eqnarray}

\noindent whereas at jump time $t_i$ the quantity is updated as
\begin{equation}
    \lim_{t \searrow t_i} \log p(y_{[0,t]}) = \lim_{t \nearrow t_i} \log p(y_{[0,t]}) + \log p_{y_{i+1}\leftarrow y_i}(t_{i+1}) \label{eq:logprob_ev_jump}
\end{equation}

\begin{minipage}{0.5\linewidth}
\begin{algorithm}[h]
\begin{algorithmic}
\State \textbf{Input:} $T$ - simulation time
\State \textbf{Output:} $y_{[0,T]}$ - trajectory
\hline 
\State \textbf{sample} $y_0 \sim p_0$
\State \textbf{set} $t = 0$, $i = 0$ 
\While{$t < T$}
\State \textbf{sample} $\tau_{i+1} \sim \mathrm{Exp}(p_{\leftarrow y_i}^{-1})$
\State \textbf{sample} $y_{i+1}$ \textbf{from} $p(\cdot \, | \, y_i)$
\State \textbf{set} $t = t + \tau_{i+1}$
\EndWhile \\
\Return states $(y_0, y_1, \ldots)$, waiting times $(\tau_0, \tau_1, \ldots)$
\end{algorithmic}
\caption{The Gillespie Algorithm, returning exact samples from a Markov chain $p$}
\label{alg:gillespie}
\end{algorithm}
\end{minipage}

\section{Kullback-Leibler Divergence between Markov Chains}

In \cite{opper_variational_nodate} the authors derive an expression for the Kullback-Leibler divergence between two continuous-time Markov chains $q$ and $p$ on an interval $[0,T]$ and a common state space $\mathcal Y$ by the following argument. For any integer $N > 1$ let $q^{(N)}$ resp.~$p^{(N)}$ be the time-discretisations of $q$ and $p$ with time step $\delta t = T / N$. These define probability distributions on $\mathcal Y^{N+1}$, with
\begin{eqnarray}
    &\KL(q^{(N)} \, \| \, p^{(N)}) =  \displaystyle\sum_{y} q_0(y) \log \frac{q_0(y)}{p_0(y)} \sum_{i=0}^{N-1} \sum_{y,y'} q^{(N)}_i(y) \, q_i^{(N)}(y' \, | \, y) \log \frac{q_i^{(N)}(y' \, | \, y)}{p_i^{(N)}(y' \, | \, y)}
\end{eqnarray}

\noindent The transition rates are related to the discrete-time transition probabilities as follows:
\begin{eqnarray}
    q^{(N)}_t(y \, | \, y) &= 1 - (\delta t) \, q_{\leftarrow y}(t) + o(\delta t) \\
    q^{(N)}_t(y' \, | \, y) &= (\delta t) \, q_{y' \leftarrow y}(t) + o(\delta t) &\qquad  (y' \neq y) 
\end{eqnarray}

\noindent In the limit $N \rightarrow \infty$ the KL divergence between the time discretisations of $q$ and $p$ converges to Eq.~\ref{eq:kl_ct}.

As in the discrete-time case the KL divergence can be written as follows in terms of the entropy $H(q)$ and the cross-entropy $H(q; p)$:
\begin{equation}
    \KL(q \,\|\, p) = H(q; p) - H(q) \label{eq:kl_diff_ent}
\end{equation}

\noindent with $H(q) = H(q; q)$ and the cross-entropy $H(q;p)$ defined as
\begin{equation}
    H(q; p) = H(q_0; p_0) + \displaystyle \int_0^T \dif t \sum_{y} q_{t}(y) p_{\leftarrow y}(t) - \int_0^T \dif t \sum_{y\neq y'} q_{t}(y)\, q_{y' \leftarrow y}(t) \log p_{y' \leftarrow y}(t)  \label{eq:crossent_ct}
\end{equation}

\noindent Starting from the discrete-time case it can similarly be shown that 
\begin{align}
    H(q; p) &= -\E_{y_{[0,T]} \sim q} \left[ \log p(y_{[0,T]}) \right] \label{eq:cent_logprob_ct}
\end{align}

The above derivation is valid only if $q_{y'\leftarrow y} \neq 0$ implies $p_{y'\leftarrow y} \neq 0$; otherwise the KL is defined to be $\infty$. The validity condition implies that a trajectory sampled from $q$ has nonzero probability under $p$ almost surely.

\section{Unobserved Variables}

We now turn to the problem of computing the likelihood of a trajectory where some variables are not observed. Consider a trajectory $x_{[0,T]}$ decomposing into an observed component $y_{[0,T]}$ and an unobserved component $z_{[0,T]}$. The jumps in the original trajectory split into two types: those changing the observed variable $y$, which we will call visible jumps, and those that only change $z$, which we will call hidden jumps. 

Again we consider the discrete-time case first, starting with the discretisation $x^{(N)}_{1:N}$ of the trajectory with time step $\delta t = T/N$, as well as its observed and unobserved components. Dropping the superscript for simplicity, the marginal likelihood of the observed trajectory $y_{1:N}$ can be computed using the forward algorithm for Hidden Markov Models\cite{} using the following recurrence relation:
\begin{equation}
    P(y_{0:i+1}, z_{i+1} = z) &= \sum_{z'} P(y_{i+1}, z \, | \, y_i, z_i = z') p(y_{0:i}, z_i = z')~.
\end{equation}

\noindent Starting with knowledge of the initial distribution $p_0$ we can then compute $p(y_{1:N},z_N = z)$ and obtain the marginal likelihood by integrating:
\begin{equation}
    P(y_{0:N}) = \sum_z P(y_{0:N}, z_N = z).
\end{equation}

We take the continuous-time limit of the above in the two separate cases $y_i \neq y_{i+1}$ (during a visible jump) and $y_i = y_{i+1}$. In the latter we obtain
\begin{eqnarray}
    \displaystyle P(y_{0:i+1}, z_{i+1} = z) &= P(y_{i+1}, z_{i+1} = z \, | \, y_i, z_i = z)\, p(y_{0:i}, z_i = z) \nonumber \\
    &\displaystyle + \sum_{z' \neq z} P(y_{i+1}, z_{i+1} = z \, | \, y_i, z_i = z') P(y_{0:i}, z_i=z').
\end{eqnarray}

\noindent As $\delta t \rightarrow 0$ this converges to
\begin{equation}
    \displaystyle P(y_{0:i+1}, z_{i+1} = z) = (1 - (\delta t) p_{\leftarrow (y_i, z)}) \, P(y_{0:i}, z_i = z) + (\delta t) \sum_{z' \neq z} p_{(y_i,z) \leftarrow (y_i, z')} \, P(y_{0:i}, z_i = z') + o(\delta t).
\end{equation}

\noindent In the limit $\delta t \rightarrow 0$ we derive the following evolution equation for the quantity $p(y_{[0,t]}, z_t = z)$:
\begin{equation}
    \frac{\dif}{\dif t} P(y_{[0,t]}, z_t = z) = -p_{\leftarrow (y_i, z)} \, P(y_{[0,t]}, z_t = z) + \sum_{z'\neq z} p_{(y_t,z) \leftarrow (y_t, z')} \, P(y_{[0,t]}, z_t = z'). \label{eq:logprob_ev_cont_latent}
\end{equation}

\noindent This resembles the Chemical Master Equation for the evolution of the probability distribution over the latent states $z$ given the observed state $y_i$. The difference to the (conditional) CME is that the first term includes all jumps out of the state $(y_i,z)$, possibly to states with $y \neq y_i$. Due to these jumps the total probability summed over all $z$ decreases with time instead of being conserved, and the remaining total probability is precisely the marginal likelihood we want to compute.

In the case of a visible jump, ie. $y_{i+1} \neq y_i$, 
\begin{equation}
    \displaystyle P(y_{0:i+1}, z_{i+1} = z) = (\delta t) \sum_{z'} p_{(y_{i+1},z) \leftarrow (y_i, z')} \, P(y_{0:i}, z_i = z') + o(\delta t)
\end{equation}

\noindent which in the continuous limit yields the following update at visible transitions:
\begin{eqnarray}
    \displaystyle  \lim_{t \searrow t_i} P(y_{[0,t]}, z_t = z) = \sum_{z'} p_{(y_{i+1},z) \leftarrow (y_i, z')}(t_i) \, \cdot \left(\lim_{t \nearrow t_i} P(y_{[0,t]}, z_t = z')\right) \label{eq:logprob_ev_jump_latent}
\end{eqnarray}

Together Eqs.~(\ref{eq:logprob_ev_cont_latent}) and (\ref{eq:logprob_ev_jump_latent}) generalise (\ref{eq:logprob_ev_cont}) and (\ref{eq:logprob_ev_jump}) to the case where not all variables are observed.

The above system of equations can be solved numerically by standard ODE integration methods. If the latent state space is infinite, which is often the case, it first has to be truncated analogously to the standard FSP with probabilities outside the truncated states set to be $0$. The truncated set of equations is a lower bound for the true marginal likelihood $p(y_{[0,T]})$. In order to optimise the marginal likelihood with respect to the model parameters $\theta$ standard autodifferentiation methods can be applied to compute gradients $\nabla_\theta p(y_{[0,T]})$.

